\documentclass[a4paper,12pt]{article}

\usepackage{amssymb,amsmath}
\usepackage{mathrsfs}
\usepackage{bbm}
\usepackage{graphicx,subfigure}

\newlength{\dinwidth}
\newlength{\dinmargin}
\begin{document}

\title{\bf \Large Anomalous $tqZ$ coupling effects in rare B- and K-meson decays}
\bigskip

\author{Xin-Qiang Li$^{1,2}$, Ya-Dong Yang$^{3,4}$ and Xing-Bo Yuan$^{3}$\\
{ $^1$\small Department of Physics, Henan Normal University, Xinxiang, Henan 453007, P.~R. China}\\
{ $^2$\small IFIC, Universitat de Val\`encia-CSIC, Apt. Correus 22085, E-46071 Val\`encia, Spain}\\
{ $^3$\small Institute of Particle Physics, Huazhong Normal University, Wuhan, Hubei 430079, P.~R. China}\\
{ $^4$\small Key Laboratory of Quark \& Lepton Physics, Ministry of Education, Huazhong Normal University}\\
{\small Wuhan, Hubei, 430079, P.~R. China}}

\date{}
\maketitle
\bigskip\bigskip
\maketitle
\vspace{-1.5cm}

\begin{abstract}
{\noindent}
As a top-factory, the LHC is performing a direct study of top-quark anomalous FCNC couplings, which are, however, correlated closely with the rare B- and K-meson decays. In this paper, we study the effects of anomalous $tqZ$~(with $q=u,c$) couplings in the rare decays $B_{s,d}\to \mu^+\mu^-$, $B\to X_s \nu \bar\nu$, $B\to K^{(*)}\nu \bar\nu$, $K^+\to \pi^+ \nu \bar\nu$, and $K_L\to \pi^0 \nu \bar\nu$. With the up-to-date experimental bounds on the branching ratios of these channels, constraints on the left-handed anomalous couplings $X_{ct}^L$ and $X_{ut}^L$ are derived, respectively. With these low-energy constraints taken into account, we find that, for real couplings $X_{ct}^L$ and $X_{ut}^L$, the indirect upper bounds on $\mathcal B(t\to qZ)$ are much lower than that from the D0 collaboration, but are still compatible with the $5\sigma$ discovery potential of ATLAS with an integrated luminosity of $10~{\rm fb}^{-1}$. With refined measurements to be available at the LHCb, the future super-B factories, the NA62 at CERN, and the KOTO at J-PARC, closer correlations between the $t\to qZ$ and the rare B- and K-meson decays are expected in the near future, which will be helpful for the searches of the top-quark FCNC decays at the LHC.

\end{abstract}

\newpage

\section{Introduction}
\label{sec:intro}

In the Standard Model~(SM), the flavor-changing neutral current~(FCNC) interactions, which are absent at tree level, are induced by quantum corrections and highly suppressed due to the Glashow-Iliopoulos-Maiani~(GIM) mechanism~\cite{Glashow:1970gm}. Possible new physics~(NP) beyond the SM can manifest itself by altering the expected rates of these FCNC-induced processes. Thus, the study of FCNC interactions plays an important role in testing the SM and probing NP effects.

For the top quark, the FCNC-induced decays $t \to q Z$~(where $q$ denotes either a $c$- or a $u$-flavored quark) are exceedingly rare within the SM, with branching ratios of order of $10^{-10}$~\cite{Eilam:1990zc,AguilarSaavedra:2004wm}. However, these processes could be significantly enhanced by some potential NP models~\cite{Beneke:2000hk}, like supersymmetry, multi-Higgs doublet models and SM extensions with exotic quarks. Any positive signal of these processes at the LHC would therefore imply NP beyond the SM. These top-quark anomalous couplings could also be probed by studying the top-quark production at high-energy colliders~\cite{Beneke:2000hk}. So far, the direct experimental bounds on these anomalous couplings are not so restrictive and the world's best limit is set by the D0 collaboration, with a branching ratio $\mathcal B(t \to qZ)<3.2\%$ at $95\%$ C.L.~\cite{Abazov:2011qf}. The constraint will be improved significantly by the large top-quark sample to be available at the CERN Large Hadron Collider~(LHC), which is expected to produce 80 million top pairs and 34 million single tops annually. For example, with about $10~{\rm fb}^{-1}$ data, the discovery potential of $\mathcal B(t\to qZ)$ at both the ATLAS~\cite{Carvalho:2007yi} and the CMS~\cite{Benucci:2008zz} collaboration is reported to be of the order of $10^{-4}$.

However, if the top-quark anomalous couplings really existed, the low-energy processes with loops involving the top quark may also be affected, and could therefore provide helpful information for a direct search at high-energy colliders~\cite{FCNC-top,arXiv:0802.1413,Han:1995pk,Yuan:2010vk,arXiv:1105.0364}. In this respect, the rare B- and K-meson decays, such as $B_{s,d}\to \mu^+\mu^-$, $B\to X_s \nu \bar\nu$, $B\to K^{(*)}\nu \bar\nu$, $K^+\to \pi^+ \nu \bar\nu$ and $K_L\to \pi^0 \nu \bar\nu$, are particularly interesting. They are all short-distance dominated FCNC processes induced by the Z-penguin and box diagrams, and the calculation of their branching ratios is theoretically rather clean. They are therefore known to be good probes of flavour dynamics within the SM and beyond~\cite{Buras:2009if}. As the anomalous couplings $tqZ$ can enter the Z-penguin diagram, constraints on these couplings can be obtained by studying deviations from the SM predictions for these decays. Although the current experimental upper bounds on these rare decay processes are still weak~\cite{Asner:2010qj}, the measurements will be improved at the LHCb, the future super-B factories, the NA62 at CERN, the KOTO at J-PARC, etc. Thus, closer correlations between the FCNC transition $t\to qZ$ and the rare B- and K-meson decays are expected in the near future.

In our previous works~\cite{Yuan:2010vk,arXiv:1105.0364}, we have studied the anomalous $tq\gamma$ coupling effects in inclusive and exclusive radiative B-meson decays. In this paper, we shall continue to investigate the anomalous $tqZ$ coupling effects on the rare B- and K-meson decays. With the up-to-date experimental data on these decays, we shall first derive constraints on these anomalous couplings, and then discuss the implications for the rare $t \to qZ$ decays at the LHC.

Our paper is organized as follows. In Sec.~\ref{sec-tqZ}, we introduce the effective Lagrangian describing the anomalous interactions $tqZ$, and set the convention used throughout the paper. In Sec.~\ref{sec-theo}, we first recapitulate the basic theoretical formulae for the relevant B- and K-meson decays, and then discuss the anomalous $tqZ$ coupling effects in these decays; the rare decay $t\to qZ$ mediated by the anomalous coupling is also presented in this section. Detailed numerical results and discussions are presented in Sec.~\ref{sec-analysis}. Our conclusions are made in Sec.~\ref{sec-conclusion}. The relevant input parameters are collected in the appendix.

\section{Effective Lagrangian for anomalous $tqZ$ couplings}
\label{sec-tqZ}

In most extensions of the SM, the new degrees of freedom that modify the ultraviolet behavior of the underlying theory appear only at a scale $\Lambda$ which is much higher than the electroweak scale $v=246~{\rm GeV}$. As long as we are only interested in processes occurring much below the scale $\Lambda$, we can always integrate out these new degrees of freedom and describe the NP effects in terms of a few higher-dimensional local operators, which are built out of the SM fields and suppressed by inverse powers of the NP scale $\Lambda$~\cite{Appelquist:1974tg,Grzadkowski:2010es,Buchmuller:1985jz,AguilarSaavedra:2008zc}.

The above effective field theory approach is a powerful theoretical framework for describing the FCNC processes induced by some unknown NP models. Specific to our case, the anomalous coupling $tqZ$ mediating the FCNC transition $t\to qZ$ can be described by the effective Lagrangian~\cite{Grzadkowski:2010es,Buchmuller:1985jz}
\begin{align}
\mathcal L^{\rm eff}_{tqZ}=\mathcal L^{\rm SM}+\sum_i \frac{C_i^{(6)}}{\Lambda^2}\mathcal O_i^{(6)}+\dots\,,
\end{align}
where operators with dimension larger than 6 are neglected. The explicit form of the dimension-6 operators $\mathcal O_i^{(6)}$, which are consistent with the SM gauge symmetries, can be found in Ref.~\cite{Grzadkowski:2010es,Buchmuller:1985jz,AguilarSaavedra:2008zc}. These operators can contribute to the $tqZ$ vertex, resulting an equivalent description by the effective Lagrangian~\cite{AguilarSaavedra:2008zc}
\begin{align}\label{Lagrangian-tqZ}
\mathcal L_{tqZ}&=\frac{g}{2\cos\theta_W}\,\bar q\gamma^\mu(X_{qt}^LP_L+X_{qt}^RP_R)t Z_\mu\nonumber\\
&\quad +\frac{g}{2\cos\theta_W}\,\bar q\frac{i\sigma^{\mu\nu}p_\nu}{M_Z}(\kappa_{qt}^LP_L+\kappa_{qt}^RP_R)\,tZ_\mu+{\rm h.c.}\,,
\end{align}
with $P_{L,R}\equiv (1\mp\gamma^5)/2$. The dimensionless couplings $X_{qt}^{L,R}$ and $\kappa_{qt}^{L,R}$ depend on the unknown Wilson coefficients $C_i^{(6)}$, and are in general complex. The effective Lagrangian given by Eq.~(\ref{Lagrangian-tqZ}) is commonly employed in phenomenological analyses related to top-quark physics~\cite{Beneke:2000hk}.

\section{Theoretical formalism for rare B- and K-meson decays}
\label{sec-theo}

In this section, we shall first recapitulate the basic theoretical formulae for the relevant rare B- and K-meson decays, and then discuss the anomalous $tqZ$ coupling effects on these decays; the rare decay $t\to qZ$ mediated by the anomalous coupling is also presented in this section.

\subsection{$B_{s,d}\to \mu^+\mu^-$}
\label{sec-bsmumu}

The rare decays $B_{s,d}\to \mu^+\mu^-$ are dominated by the Z-penguin and box diagrams involving top-quark exchanges, and the resulting effective Hamiltonian can be written as~\cite{hep-ph/9901288,hep-ph/9901278}
\begin{align}\label{Hamiltonian-b2smumu}
\mathcal H_{\rm eff}&=-\frac{G_{\rm F}}{\sqrt 2}\frac{\alpha}{2\pi\sin^2\theta_W}V_{tb}^*V_{tq}Y(x_t)(\bar b q)_{V-A}(\bar \mu \mu)_{V-A}+{\rm h.c.}\,,
\end{align}
where $(\bar f f^{\prime})_{V-A}\equiv \bar f \gamma^\mu(1-\gamma^5)f^{\prime}$, $q=s(d)$ for $B_s(B_d)$-meson decay, and the gauge-invariant function $Y(x_t)$ is a linear combination of the $V-A$ components of Z-penguin and box diagrams~\cite{hep-ph/9901288,hep-ph/9901278,Inami:1980fz}, with its explicit expression given in Appendix~B. The branching ratio for $B_q \to \mu^+\mu^-$ is then given by~\cite{hep-ph/9901288,hep-ph/9901278}
\begin{align}
\mathcal B(B_q\to \mu^+\mu^-)&=\frac{G_{\rm F}^2}{\pi}\Bigl(\frac{\alpha}{4\pi\sin^2\theta_W}\Bigr)^2 |V_{tb}^*V_{tq}|^2 f_{B_q}^2 m_\mu^2 m_{B_q} \tau_{B_q} \sqrt {1-\frac{4m_\mu^2}{m^2_{B_s}}}\, |Y(x_t)|^2\,,
\end{align}
where $f_{B_q}$ is the $B_q$-meson decay constant. Due to the helicity suppression factor $m_\mu^2$, the branching ratios for these decays are predicted to be very small.

\subsection{Exclusive and inclusive $b\to s\nu\bar\nu$ decays}
\label{sec-bsnunu}

The rare decays $B\to X_s\nu\bar\nu$, $B\to K \nu\bar\nu$ and $B\to K^* \nu\bar\nu$ are all induced by the quark-level $b\to s \nu\bar\nu$ transition, and provide a very good test of modified Z-penguin contributions~\cite{Altmannshofer:2009ma,Bartsch:2009qp,Colangelo:1996ay}. The effective weak Hamiltonian governing the transition $b\to s\nu\bar\nu$ can be written as~\cite{hep-ph/9901288,hep-ph/9901278}
\begin{align} \label{Hamiltonian-b2snunu}
\mathcal H_{\rm eff}=\frac{G_F}{\sqrt 2}\frac{\alpha}{2\pi\sin^2\theta_W}V_{tb}V_{ts}^* X(x_t)(\bar s b)_{V-A}(\bar\nu \nu)_{V-A}+{\rm h.c.}\,,
\end{align}
where the gauge-invariant function $X(x_t)$ is also a linear combination of the $V-A$ components of Z-penguin and box diagrams~\cite{hep-ph/9901288,hep-ph/9901278,Inami:1980fz}. For convenience, we give its explicit expression in Appendix~B.

\subsubsection{$B \to X_s \nu\bar\nu$}

The inclusive decay $B\to X_s\nu\bar\nu$ can be evaluated using heavy-quark expansion and operator product expansion, and is theoretically very clean. Adopting the convention advocated by Ref.~\cite{Altmannshofer:2009ma}, the dineutrino invariant mass distribution can be written as
\begin{align}\label{eq:B2Xsnunubar}
\frac{d\Gamma (B \to X_s\nu\bar\nu)}{ds_b}=&\frac{G_F^2\alpha^2m_b^5}{128\pi^5\sin^4\theta_W}
|V_{tb}V_{ts}^*|^2\kappa(0)|X(x_t)|^2 \nonumber\\
&\qquad \qquad \times\sqrt{\lambda(1,\hat m_s^2,s_b)}\, \left[3s_b\left(1+\hat m_s^2-s_b\right)+\lambda(1,\hat m_s^2,s_b)\right]\,,
\end{align}
with $\lambda(x,y,z)=x^2+y^2+z^2-2(xy+yz+zx)$, $\hat m_i=m_i/m_b$, and $m_b$ denotes the $b$-quark mass in the 1S scheme~\cite{Hoang:1998ng,Bauer:2004ve}. The factor $\kappa(0)=0.83$ contains the virtual and bremsstrahlung QCD corrections to the $b\to s\nu\bar\nu$ matrix element~\cite{Buchalla:1993bv,Grossman:1995gt}. The total branching ratio is then obtained by integrating Eq.~(\ref{eq:B2Xsnunubar}) over the kinematically allowed region $0\le s_b=q^2/m_b^2 \le (1-\hat m_s)^2$, where $q^2$ is the invariant mass of the neutrino-antineutrino pair. In addition, we have also incuded the additional $\mathcal O(\Lambda^2/m_b^2)$ corrections~\cite{Grossman:1995gt,Falk:1995me}, with the HQET parameters $\lambda_1=-0.27\pm 0.04$ and $\lambda_2=0.12\pm 0.01$~\cite{Bauer:2004ve}.

\subsubsection{$B \to K \nu\bar\nu$}

For the exclusive decay $B\to K\nu\bar\nu$, the dineutrino invariant mass distribution can be written as~\cite{Altmannshofer:2009ma,Colangelo:1996ay}
\begin{align}
\frac{d\Gamma(B\to K\nu\bar\nu)}{ds_B}=\frac{G_F^2\alpha^2m_B^5}{256\pi^5\sin^4\theta_W}\, |V_{tb}V_{ts}^*|^2\, \lambda^{3/2}(s_B,\widetilde m_K^2,1)\,[f_+^K(s_B)]^2\, |X(x_t)|^2,
\end{align}
where $\widetilde{m}_K=m_K/m_B$, and $s_B=q^2/m_B^2$ is constrained within the physical region $0 \le s_B  \le (1 -\widetilde{m}_K )^2 \approx 0.82$; $f_+^K(s_B)$ is the $B\to K$ transition form factor, the presence of which results in large theoretical uncertainty and makes the exclusive mode not as clean as the inclusive one. However, significant progress has recently been made by considering simultaneously also the decay mode $B\to K \mu^+ \mu^-$~\cite{Bartsch:2009qp}.

\subsubsection{$B\to K^*\nu\bar\nu$}

For the decay $B\to K^*\nu\bar\nu$, additional information about the polarization of the $K^*$ meson can be extracted from the angular distribution of the $K^*$ decay products. In terms of the three transversity amplitudes~\cite{Altmannshofer:2009ma}
\begin{align}
A_\perp(s_B)&=-2N \sqrt 2 \lambda^{1/2}(1,\widetilde m _{K^*}^2,s_B) X(x_t) \frac{V(s_B)}{(1+\widetilde m_{K^*})}\,, \nonumber\\
A_\parallel(s_B)&=2N \sqrt 2 (1+\widetilde m_{k^*}) X(x_t) A_1(s_B)\,, \nonumber\\
A_0(s_B)&=\frac{N X(x_t)}{\widetilde m_{K^*}\sqrt {s_B}}\left[(1-\widetilde m_{K^*}^2-s_B) (1+\widetilde m_{K^*}) A_1(s_B)-\lambda(1,\widetilde m_{K^*}^2,s_B) \frac{A_2(s_B)}{1+\widetilde m_{K^*}}\right]\,,
\end{align}
with an overall factor
\begin{align}
N = V_{tb}V_{ts}^* \left[\frac{G_F^2 \alpha^2 m_B^3}{3 \cdot 2^{10} \pi^5\sin^4\theta_W} s_B \lambda^{1/2}(1,\widetilde m_{K^*}^2,s_B)\right]^{1/2},
\end{align}
the dineutrino invariant mass spectrum of the decay $B\to K^*\nu\bar\nu$ can be written as~\cite{Altmannshofer:2009ma}
\begin{align}
\frac{d\Gamma}{ds_B}=3m_B^2(|A_\perp(s_B)|^2+|A_\parallel(s_B)|^2+|A_0(s_B)|^2)\,,
\end{align}
where the normalized invariant mass $s_B$ ranges from 0 to the kinematical endpoint $(1-\widetilde{m}_{K^*})^2\approx0.69$. Here the main theoretical uncertainty is due to the normalization and the shape of the three $B\to K^*$ transition form factors $V(q^2)$, $A_1(q^2)$ and $A_2(q^2)$.

\subsection{Rare K-meson decays}
\label{sec-sdnunu}

It is known that, among the many rare B- and K-meson decays, the modes $K^+\to \pi^+\nu\bar\nu$ and $K_L\to \pi^0\nu\bar\nu$ are the theoretically cleanest, and therefore play an important role in the search for the underlying mechanism of flavour mixing and CP violation~\cite{hep-ph/0405132,arXiv:1107.6001}. Especially, these decays are very sensitive to NP contributions in Z-penguin diagrams.

\subsubsection{$K^+\to \pi^+\nu\bar\nu$}

The effective weak Hamiltonian relevant for $K^+\to \pi^+ \nu\bar\nu$ can be written as~\cite{hep-ph/9901288,hep-ph/9308272}
\begin{align}
\mathcal H_{\rm eff}=\frac{G_{\rm F}}{\sqrt 2}\frac{\alpha}{2 \pi \sin^2\theta_W} \sum_{l=e,\mu,\tau} \Bigl(V_{cs}^*V_{cd} X_{NL}^l + V_{ts}^*V_{td} X(x_t)\Bigl)(\bar s d)_{V-A}(\bar\nu_l\nu_l)_{V-A}\,,
\end{align}
where the dependence on the charged lepton mass results from the box diagram. With the help of isospin symmetry, the branching ratio for $K^+\to\pi^+\nu\bar\nu$ is then given as~\cite{hep-ph/0405132,arXiv:1107.6001}
\begin{align}
\mathcal B(K^+\to\pi^+\nu\bar\nu)&=\kappa_+ (1+\Delta_{\rm EM}) \left|\frac{\lambda_t} {\lambda^5} X(x_t) +\frac{\lambda_c} {\lambda}(P_c+\delta P_{c,u})\right|^2,
\end{align}
where $\lambda_q=V_{qs}^*V_{qd}$~(with $q=t,c$), is the product of the Cabibbo-Kobayashi-Maskawa~(CKM) matrix elements~\cite{CKM}, and $\Delta_{\rm EM}=-0.003$ denotes the electromagnetic correction~\cite{arXiv:0705.2025}. The overall factor $\kappa_+$, with $\kappa_+=(5.173 \pm 0.025) \times (\lambda/0.225)^8 \times 10^{-11}$, summarizes the isospin-breaking corrections in relating $K^+\to\pi^+\nu\bar\nu$ to $K^+\to\pi^0 \ell^+ \nu_{\ell}$~\cite{arXiv:0705.2025,428572}. The dimension-six charm operator contribution $P_c$ is given as~\cite{hep-ph/0508165}
\begin{align}
P_c=\frac{1}{\lambda^4} \left(\frac{2}{3} X_{NL}^e + \frac{1}{3}X_{NL}^\tau\right)=0.38\pm 0.04\,,
\end{align}
with error dominated by the charm-quark mass, and $\delta P_{c,u}=0.04 \pm 0.02$ contains the small long-distance~(up quark) and dimension-eight charm operator contributions~\cite{hep-ph/0503107}.

\subsubsection{$K_L\to \pi^0\nu\bar\nu$}

The rare decay $K_L\to\pi^0\nu\bar\nu$, proceeding almost entirely through direct CP violation~\cite{BNL-42227}, is completely dominated by short-distance loop diagrams with top-quark exchanges, and the charm contribution can be fully neglected~\cite{hep-ph/0405132}. Thus, the effective weak Hamiltonian for $K_L\to\pi^0\nu\bar\nu$ can be written as~\cite{hep-ph/9901288,hep-ph/9308272}
\begin{align}
\mathcal H_{\rm eff}=\frac{G_F}{\sqrt 2}\frac{\alpha}{2\pi\sin^2\theta_W}V_{ts}^*V_{td}X(x_t)(\bar s d)_{V-A}(\bar \nu \nu)_{V-A}+{\rm h.c.}\,.
\end{align}
Analogous to the case of $K^+\to\pi^+\nu\bar\nu$, the branching ratio of $K_L\to\pi^0\nu\bar\nu$ can be given as~\cite{hep-ph/0405132,hep-ph/9308272}
\begin{align}
\mathcal B(K_L\to\pi^0\nu\bar\nu)= \kappa_L \left(\frac{{\rm Im}[\lambda_t X(x_t)]} {\lambda^5}\right)^2(1-\delta_\epsilon)\,,
\end{align}
where the overall factor $\kappa_L=(2.231 \pm 0.013)\times (\lambda/0.225)^8 \times 10^{-10}$ encodes the hadronic matrix element related to $K_{\ell3}$ data~\cite{arXiv:0705.2025,428572}, and the parameter $\delta_\epsilon$ denotes the contribution of indirect CP violation to $K_L\to\pi^0\nu\bar\nu$, which is highly suppressed by the small $K^0-\bar{K}^0$ mixing parameter $|\epsilon|$~\cite{hep-ph/9308272}.
For the case of general complex function $X(x_t)$, the expression of $\delta_\epsilon$ can be written as
\begin{align}
\delta_\epsilon =-\sqrt 2 |\epsilon| \frac{{\rm Re}[\lambda_t X(x_t)]+{\rm Re}(\lambda_c) \lambda^4 P_c}{{\rm Im}[\lambda_t X(x_t)]}\,,
\end{align}
which is consistent with the SM result for real $X(x_t)$ given explicitly in Ref.~\cite{hep-ph/9308272}.

\subsection{Anomalous $tqZ$ coupling effects}
\label{sec-tqzeffect}

The anomalous $tqZ$ interactions given by Eq.~(\ref{Lagrangian-tqZ}) affect the rare B- and K-meson decays through the Z-penguin diagrams. As an illustration, in the following we shall consider the effect of anomalous $tqZ$ couplings on the $b\to s\nu\bar\nu$ transition.

\subsubsection{$b\to s\nu\bar\nu$ transition induced by anomalous $tcZ$ coupling}

\begin{figure}[t]
\centering
\subfigure[]{\includegraphics[width=4cm]{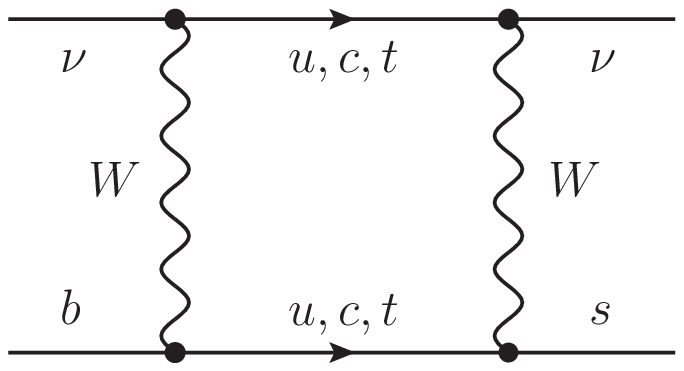}\label{b2snunu1}}
\subfigure[]{\includegraphics[width=4cm]{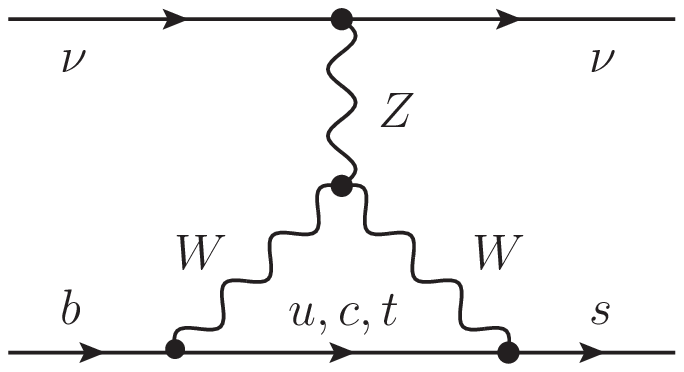}\label{b2snunu2}}
\subfigure[]{\includegraphics[width=4cm]{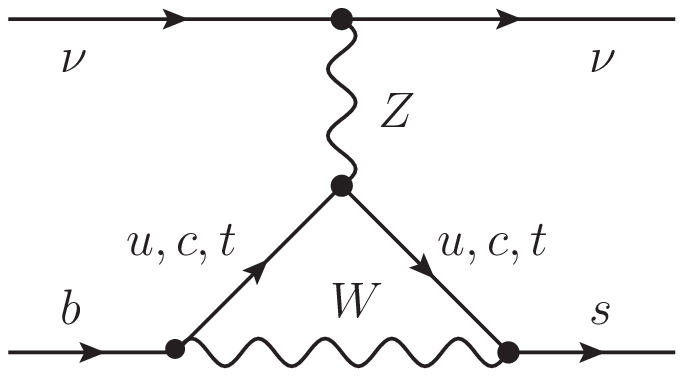}\label{b2snunu3}}
\subfigure[]{\includegraphics[width=4cm]{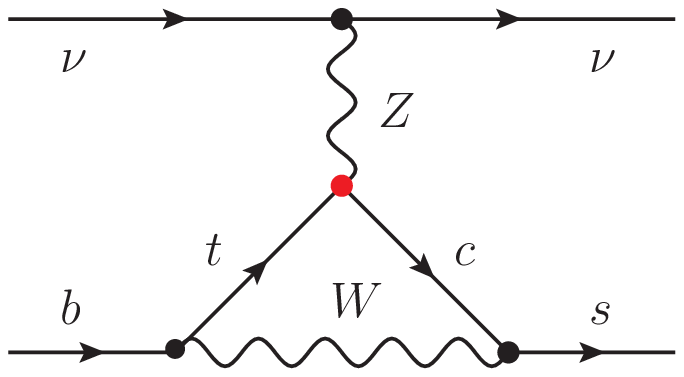}\label{b2snunuNP}}
\caption{\small Relevant Feynman diagrams for $b\to s\nu\bar\nu$ transition in the unitary gauge, where the first three ones are the one-loop SM contributions, while the last one denotes the contribution induced by the anomalous $tcZ$ coupling.}
\label{fig:b2snunu}
\end{figure}

For the transition $b\to s\nu\bar\nu$, the relevant Feynman diagrams both within the SM~(the first three ones) and with the anomalous $tcZ$ coupling~(the last one) are depicted in Fig.~\ref{fig:b2snunu}.

It should be noted that there exist some other Feynman diagrams induced by the anomalous $tqZ$ couplings, such as the one with the c- replaced by the u-quark line in Fig.~\ref{b2snunuNP}, and the one with exchanges of the t- and c-quark lines in Fig.~\ref{b2snunuNP}. However, analogous to the arguments made in our previous works~\cite{Yuan:2010vk,arXiv:1105.0364}, contributions from these two Feynman diagrams are negligible compared to the one from Fig.~\ref{b2snunuNP}, based on the observation that the associated CKM factors $|V_{cs}|>|V_{us}|$, $|V_{cb}|>|V_{ub}|$, and $|V_{tb}V_{cs}^*| \gg |V_{cb}V_{ts}^*|$. Thus, for the $b\to s\nu\bar\nu$ transition, we need only consider Fig.~\ref{b2snunuNP} with only one anomalous coupling $tcZ$.

It is also observed that the large CKM factor $|V_{tb}V_{cs}^*|\approx 1$ associated with Fig.~\ref{b2snunuNP}, compared to the SM case $|V_{tb}V_{ts}^*|\sim \mathcal O(\lambda^2)$, make the $b\to s\nu\bar\nu$ transition to be very sensitive to the anomalous couplings $tcZ$. It is therefore expected that constraint on the coupling $tcZ$ could be obtained from the precisely measured rare B-meson decays induced by the quark-level $b\to s\nu\bar\nu$ transition.

The calculation of Fig.~\ref{b2snunuNP} could be most conveniently performed in the unitary gauge, where the pseudo-Goldstone components of the SM Higgs doublet are absent. It is noted that, in the unitary gauge, the contribution from the $q^{\alpha}q^{\beta}/m_{W}^2$ part of the $W$-boson propagator is ultraviolet-divergent. Similar to the treatment adopted by Grzadkowski and Misiak~\cite{arXiv:0802.1413}, the divergence could be absorbed by some counterterms served by other dimension-6 operators~\cite{Grzadkowski:2010es,Buchmuller:1985jz}, the ${\rm \overline{MS}}$-renormalized Wilson coefficients of which are assumed to be negligible compared to the ones considered here. Moreover, it is found that the contribution from tensor currents in the effective Lagrangian Eq.~(\ref{Lagrangian-tqZ}) is zero.

Normalized to the effective Hamiltonian Eq.~(\ref{Hamiltonian-b2snunu}), the contribution of anomalous $tcZ$ coupling to the $b\to s\nu\bar\nu$ transition would result in the deviation
\begin{align}
X(x_t) \to X'=X(x_t)+C_{0,b\to s}^{\rm NP}\,,
\end{align}
where the matching coefficient reads
\begin{align}\label{X-b2s}
C_{0,\,b\to s}^{\rm NP}(\mu)=&-\frac{1}{8} \frac{V_{cs}^*}{V_{ts}^*}\biggl[X_{ct}^L \left(-x_t\log\frac{m_W^2}{\mu^2} + \frac{3}{2} + x_t - x_t\log x_t\right) \nonumber\\[0.2cm]
&\qquad\quad + X_{ct}^R \frac{\sqrt{x_{c}x_t}}{2} \left(\log\frac{m_W^2}{\mu^2}-\frac{1}{2}+\frac{x_t-4}{x_t-1}\log x_t\right) \biggr]\,,
\end{align}
with $x_q=\bar m_q^2/m_W^2$. The presence of logarithms $\ln(m_{W}^2/\mu^2)$ results from the ${\rm \overline{MS}}$ prescription for the ultraviolet-divergence. Neglecting the light charm-quark mass $m_{c}$, as done in the SM, the effect of right-handed current $X_{ct}^R$ could be safely neglected.

\subsubsection{$b(s)\to d\nu\bar\nu$ transition induced by anomalous $tuZ$ coupling}

\begin{table}[t]
\begin{center}
\caption{\label{tableCKM} \small The counting of CKM factors for $b\to sZ^*$, $b\to dZ^*$ and $s\to dZ^*$ transitions both within the SM and with the anomalous $tqZ$ couplings. Within the SM, the CKM factors in the Z-penguin and box diagrams are the same.}
\vspace{0.2cm}
\doublerulesep 0.8pt \tabcolsep 0.15in
\begin{tabular}{r l l l}
\hline\hline
& $b\to sZ^*$\,-transition & $b\to dZ^*$\,-transition & $s\to dZ^*$\,-transition \\
\hline
top-sector & $|V_{tb}V_{ts}^*|\sim \mathcal O(\lambda^2)$ & $|V_{tb}V_{td}^*|\sim \mathcal O(\lambda^3)$ & $|V_{ts}V_{td}^*|\sim \mathcal O(\lambda^5)$ \\
$tcZ$-coupling & $|V_{tb}V_{cs}^*|\sim 1$ & $|V_{tb}V_{cd}^*|\sim \mathcal O(\lambda)$ & $|V_{ts}V_{cd}^*|\sim \mathcal O(\lambda^3)$ \\
$tuZ$-coupling & $|V_{tb}V_{us}^*|\sim O(\lambda)$ & $|V_{tb}V_{ud}^*|\sim 1$ & $|V_{ts}V_{ud}^*|\sim \mathcal O(\lambda^2)$ \\
\hline \hline
\end{tabular}
\end{center}
\end{table}

Similar to the case of $b\to s\nu\bar\nu$ transition, it is easily seen that, based on the counting of CKM factors listed in Table~\ref{tableCKM}, the transitions $b\to d\nu\bar\nu$ and $s\to d\nu\bar\nu$ are both dominated by the anomalous $tuZ$ coupling, with the corresponding Feynman diagram obtained from Fig.~\ref{b2snunuNP}, with the c- replaced by the u-quark line and changes of the external quark flavours.

Normalized to the corresponding effective weak Hamiltonian, the contribution of anomalous $tuZ$ coupling to the transitions $b\to d\nu\bar\nu$ and $s\to d\nu\bar\nu$ can be written, respectively, as
\begin{align}
C_{0,\,b\to d}^{\rm NP}(\mu)=& -\frac{1}{8} \frac{V_{ud}^*}{V_{td}^*}\biggl[X_{ut}^L \left(-x_t\log\frac{m_W^2}{\mu^2} + \frac{3}{2} + x_t - x_t\log x_t\right) \nonumber \\[0.2cm]
&\qquad\quad + X_{ut}^R \frac{\sqrt{x_{u}x_t}}{2} \left(\log\frac{m_W^2}{\mu^2}-\frac{1}{2}+\frac{x_t-4}{x_t-1}\log x_t\right) \biggr]\,, \label{C0b2d} \\
C_{0,\,s\to d}^{\rm NP}(\mu)=& C_{0,\,b\to d}^{\rm NP}(\mu)\,, \label{C0s2d}
\end{align}
where the right-handed coupling $X_{ut}^R$ is now more severely suppressed by the light up-quark mass via the factor $\sqrt x_u=\bar m_u/m_W$.

\subsection{Rare $t\to qZ$ decays mediated by anomalous $tqZ$ coupling}
\label{sec-tqzdecays}

Within the SM, the top quark has only one dominant decay channel $t\to bW$, and the branching ratio of $t\to qZ$ decay can be therefore defined as~\cite{Beneke:2000hk,Han:1995pk}
\begin{align}
\mathcal B(t\to qZ)=\frac{\Gamma(t\to qZ)}{\Gamma(t\to bW)}\,,
\end{align}
where the leading-order~(LO) decay width of $t\to b W$ is given explicitly as~\cite{Li:1990qf}
\begin{align}
\Gamma_0(t \to bW) = \frac{G_{\rm F} m_t^3}{8\sqrt{2}\,\pi}|V_{tb}|^2\beta_W^4(3-2\beta_W^2)\,,
\end{align}
with $\beta_W=(1-m_W^2/m_t^2)^{1/2}$, being the velocity of the $W$-boson in the top-quark rest frame. As the SM prediction for $\Gamma(t \to q Z)$ is exceedingly small~\cite{Eilam:1990zc}, we need only consider the decay $t\to qZ$ mediated by the anomalous $tqZ$ interaction. Adopting the convention specified in the effective Lagrangian Eq.~(\ref{Lagrangian-tqZ}), the LO decay width of $t\to qZ$ can be written as~\cite{Beneke:2000hk,Han:1995pk}
\begin{align}\label{eq:t2qZ_LO}
\Gamma_0(t\to qZ) &= \frac{G_{\rm F} m_t^3}{8\sqrt{2}\,\pi}\frac{|X_{ct}^L|^2+|X_{ct}^R|^2}{2}\beta_Z^4(3-2\beta_Z^2)\,,
\end{align}
where $\beta_Z=(1-m_Z^2/m_t^2)^{1/2}$, is the velocity of the $Z$-boson in the top-quark rest frame. Since the NLO QCD corrections to $\Gamma(t\to qZ)$ are found to be negligible~\cite{Zhang:2010bm,Zhang:2008yn,Drobnak:2010by,Drobnak:2010wh}, here we shall only consider the LO result given by Eq.~(\ref{eq:t2qZ_LO}).

\section{Numerical results and discussions}
\label{sec-analysis}

With the theoretical framework presented in previous sections and the input parameters collected in Appendix~A, we shall present our numerical results and discussions in this section.

\subsection{The SM predictions and the experimental data}

\begin{table}[t]
\begin{center}
\caption{\label{table-theoexp} \small The SM predictions for the rare B- and K-meson decays, with the corresponding experimental data given in the third column. Here we give only the world's best existing limits on each decay mode with the reference given in the last column.}
\vspace{0.2cm}
\doublerulesep 0.8pt \tabcolsep 0.20in
\begin{tabular}{l l l r}
\hline\hline
Observables & SM prediction & Experimental data & Ref. \\
\hline
$\mathcal B (B_s \to \mu^+\mu^-)$ & $(3.65^{+0.32}_{-0.35}) \times 10^{-9 }$ & $<0.90 \times 10^{-8}$~(90\% C.L.) & \cite{CMSLHCb}\\
& & $<1.08 \times 10^{-8}$~(95\% C.L.) & \cite{CMSLHCb}\\
$\mathcal B (B_d \to \mu^+\mu^-)$ & $(1.08^{+0.13}_{-0.14}) \times 10^{-10}$ & $<2.6 \times
10^{-9}$~(90\% C.L.) & \cite{:2011zq}\\
& & $<3.2 \times 10^{-9}$~(95\% C.L.) & \cite{:2011zq}\\
$\mathcal B (B \to X_s \nu \bar\nu)$ & $(3.13^{+0.14}_{-0.20}) \times 10^{-5}$ & $<6.4 \times 10^{-4}$~(90\% C.L.) & \cite{hep-ex/0010022}\\
$\mathcal B (B^+ \to K^+ \nu\bar\nu)$ & $(5.29^{+0.76}_{-0.74}) \times 10^{-6}$ & $<1.3 \times
10^{-5}$~(90\% C.L.) & \cite{arXiv:1009.1529}\\
$\mathcal B (B^0 \to K^0 \nu\bar\nu)$ & $(4.91^{+0.70}_{-0.69}) \times 10^{-6}$ & $<5.6 \times
10^{-5}$~(90\% C.L.) & \cite{arXiv:1009.1529}\\
$\mathcal B (B^+ \to K^{*+} \nu\bar\nu)$ & $(1.11^{+0.25}_{-0.23}) \times 10^{-5}$ & $<8.0 \times
10^{-5}$~(90\% C.L.) & \cite{arXiv:0808.1338}\\
$\mathcal B (B^0 \to K^{*0} \nu\bar\nu)$ & $(1.03^{+0.23}_{-0.21}) \times 10^{-5}$ & $<12 \times
10^{-5}$~(90\% C.L.) & \cite{arXiv:0808.1338}\\
\hline
$\mathcal B (K^+\to\pi^+\nu\bar\nu)$ & $(8.52^{+0.70}_{-0.92}) \times 10^{-11}$ & $(1.73^{+1.15}_{-1.05}) \times 10^{-10}$ & \cite{arXiv:0808.2459}\\
$\mathcal B (K_L\to\pi^0\nu\bar\nu)$ & $(2.67^{+0.29}_{-0.36}) \times 10^{-11}$ & $<2.6 \times 10^{-8}$~(90\% C.L.) & \cite{arXiv:0911.4789}\\
\hline
$\mathcal B (t \to qZ )$ & &$<3.2\%$~(95\% C.L.) & \cite{Abazov:2011qf}\\
\hline \hline
\end{tabular}
\end{center}
\end{table}

Within the SM, our predictions for the rare B- and K-meson decays are listed in Table~\ref{table-theoexp}, where the theoretical uncertainties are obtained by varying each input parameter within its respective range and adding the individual uncertainty in quadrature. It can be seen that, with the up-to-date input parameters, the theoretical uncertainties for most of these decays are less than $13\%$ except for $B\to K^*\nu\bar\nu$ decay, which is still about $23\%$ mainly due to the $B\to K^*$ transition form factors.

However, compared to the precise theoretical predictions, the current experimental limits on these decays are still rather weak. At present, only seven events of the decay $K^+\to \pi^+\nu\bar\nu$ have been observed~\cite{arXiv:0808.2459}. For $B_s\to \mu^+ \mu^-$ decay, it is interesting to note that a possible first signal has been recently announced by the CDF collaboration, although with a
low significance~\cite{arXiv:1107.2304}. This result has, unfortunately, not been confirmed by the searches both at the CMS~\cite{CMSLHCb,arXiv:1107.5834} and at the LHCb collaboration~\cite{CMSLHCb,:2011zq,arXiv:1110.2411}. Because of the missing multiple neutrinos in the final state, it is quite difficult to measure the exclusive $b\to s \nu \bar\nu$ decays, and the present experimental limits are only available from the two $e^+ e^-$ B-factories BaBar~\cite{arXiv:1009.1529,arXiv:0808.1338} and Belle~\cite{hep-ex/0507034,arXiv:0707.0138}, both of which have used the hadronic tag technique to reconstruct the accompanying B meson.

To discuss the effects of anomalous $tqZ$ interactions on these rare B- and K-meson decays, we shall use the SM predictions with $2\sigma$ error bars and the experimental upper bounds at 90\% C.L., as listed in  Table~\ref{table-theoexp}. For the decay $K^+\to\pi^+\nu\bar\nu$, on the other hand, the experimental data with $1\sigma$ error bar will be used due to its large uncertainty. In additional, for the exclusive $b\to s\nu\bar\nu$ decays, since the experimental upper bounds on the charged decay modes, $\mathcal B(B^+\to K^+\nu\bar\nu)$ and $\mathcal B(B^+\to K^{*+}\nu\bar\nu)$, are more stringent than their neutral counterparts, $\mathcal B(B^0\to K^0\nu\bar\nu)$ and $\mathcal B(B^0\to K^{*0}\nu\bar\nu)$, we shall only consider constraints from the former in the following discussions.

For the rare top-quark FCNC decay $t\to qZ$, the current world's best limit, $\mathcal B(t\to qZ)<3.2\%$ at $95\%$ C.L., is set by the D0 collaboration~\cite{Abazov:2011qf}. On the other hand, using likelihood-based analyses, the expected branching ratio sensitivity for a $5\sigma$ discovery potential at the LHC could reach $4.4~(1.4)\times 10^{-4}$ with an integrated luminosity of $L=10~(100)~{\rm fb}^{-1}$ at the ATLAS experiment~\cite{Carvalho:2007yi}.

\subsection{$B_s\to\mu^+\mu^-$ and $b\to s\nu\bar\nu$ decays with anomalous coupling $X_{ct}^L$}

\begin{figure}[htbp]
\centering
 \subfigure{\includegraphics [width=7cm]{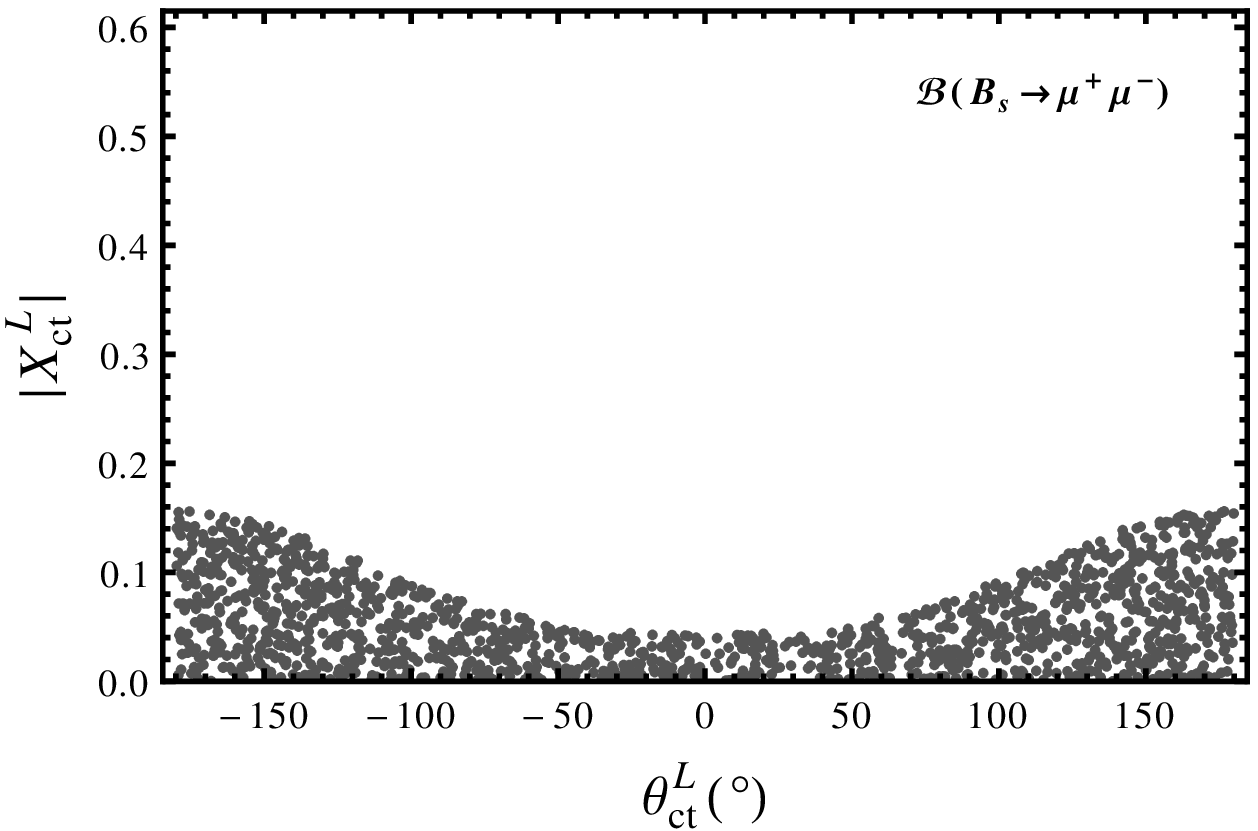}}\hspace{0.5in}
 \subfigure{\includegraphics [width=7cm]{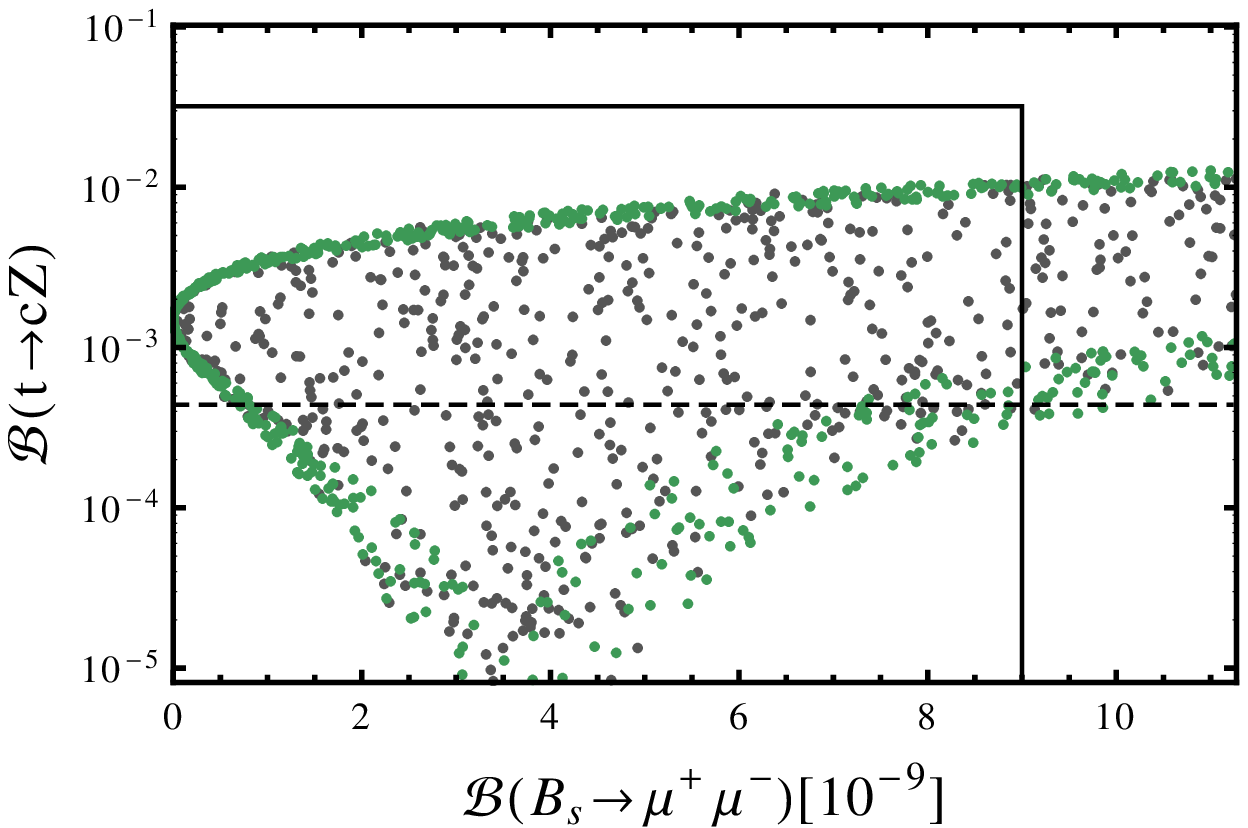}}
 \subfigure{\includegraphics [width=7cm]{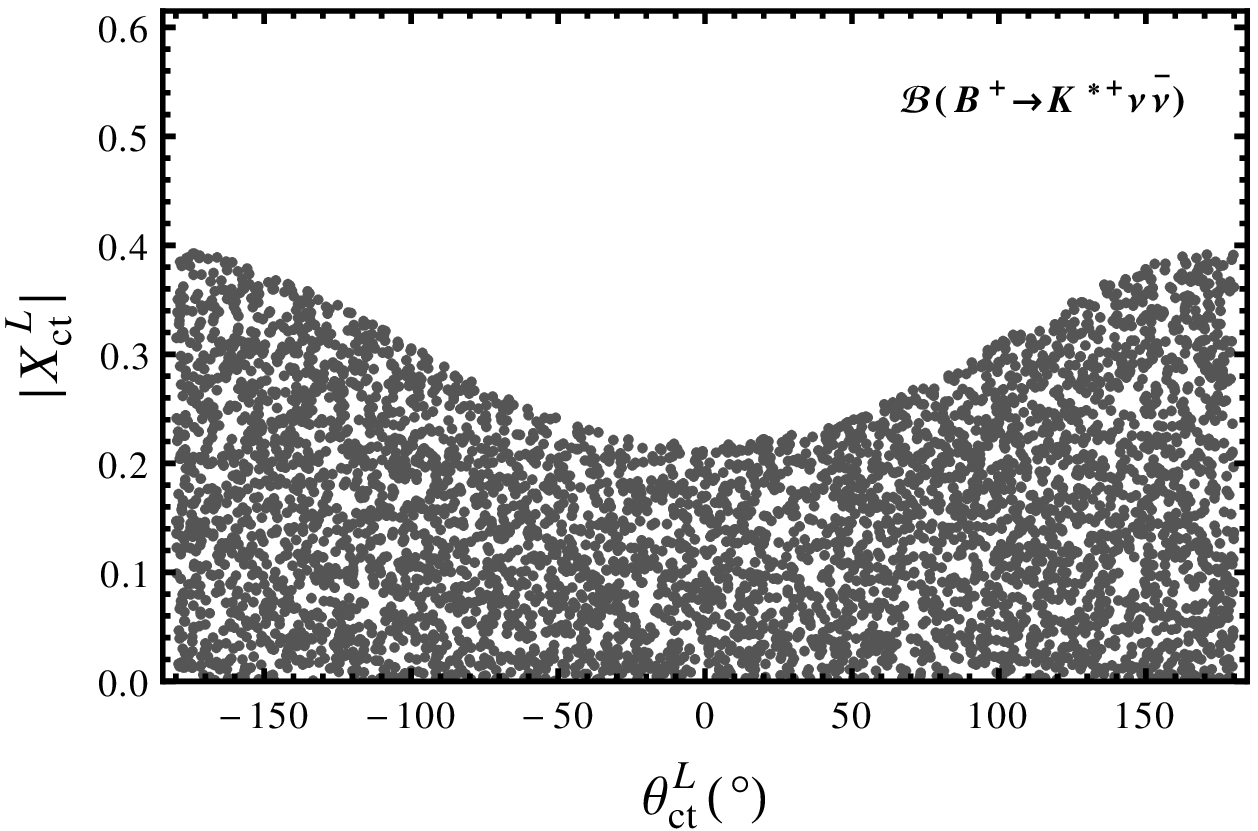}}\hspace{0.5in}
 \subfigure{\includegraphics [width=7cm]{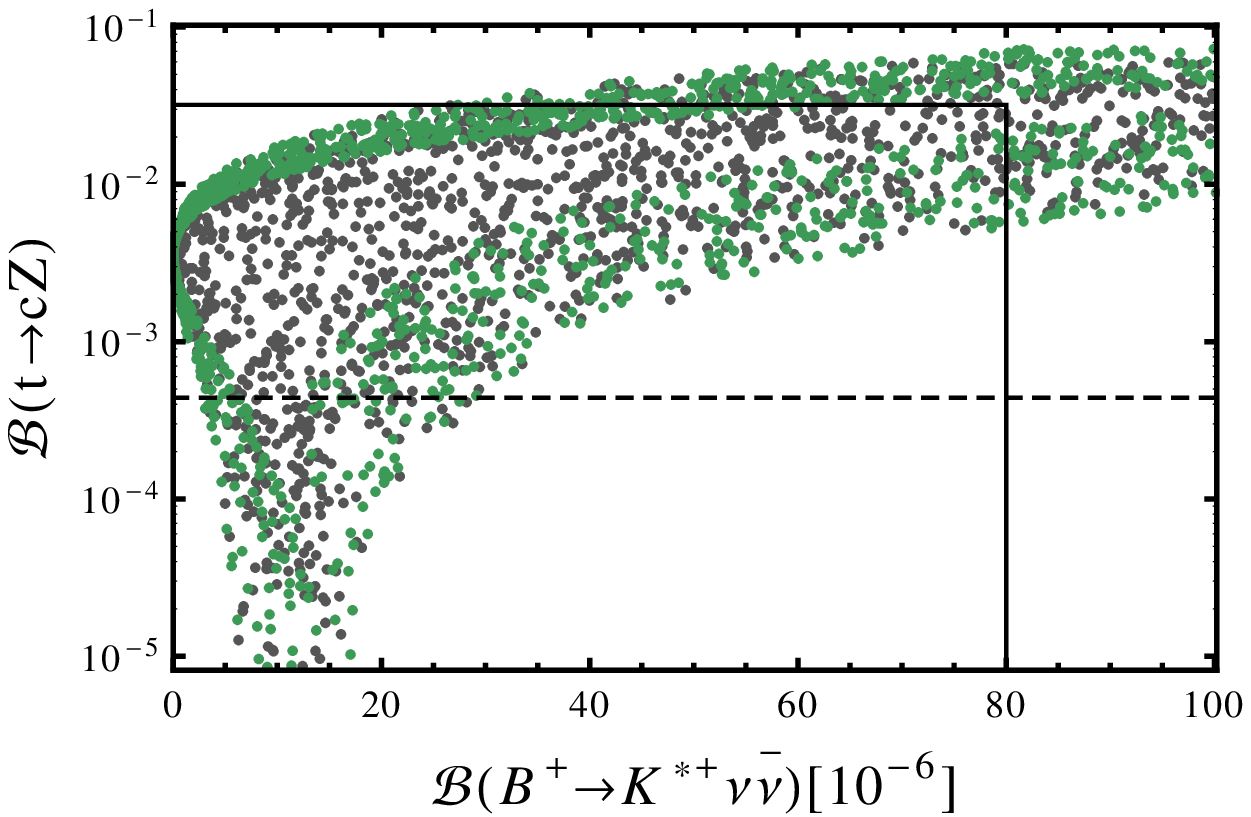}}
 \subfigure{\includegraphics [width=7cm]{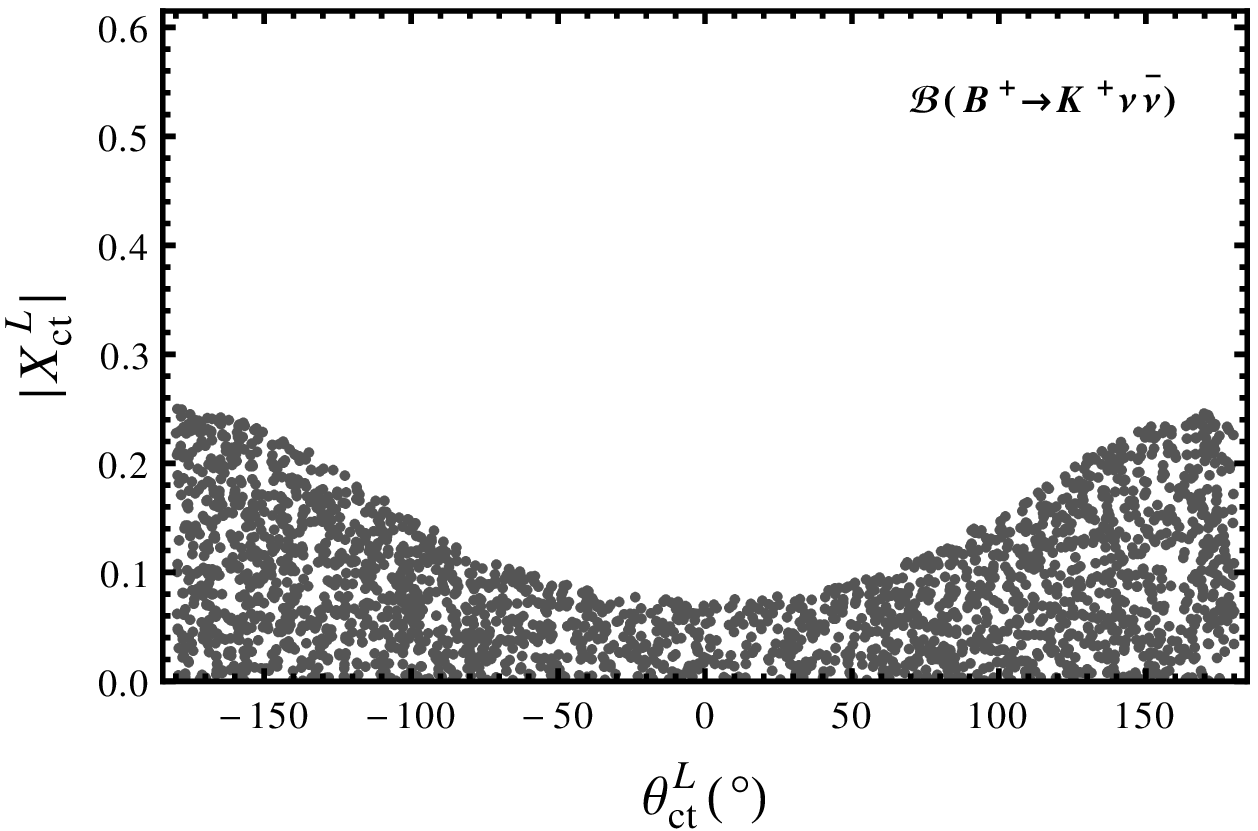}}\hspace{0.5in}
 \subfigure{\includegraphics [width=7cm]{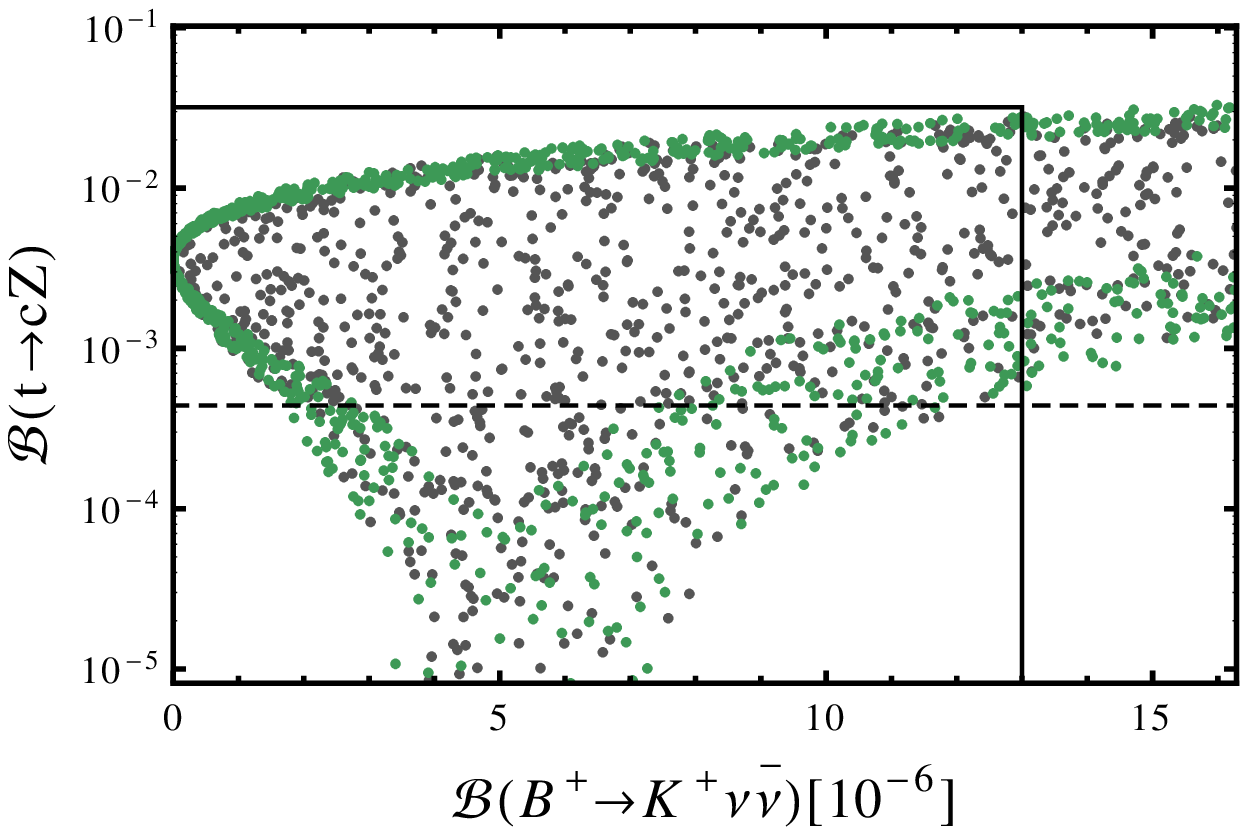}}
 \subfigure{\includegraphics [width=7cm]{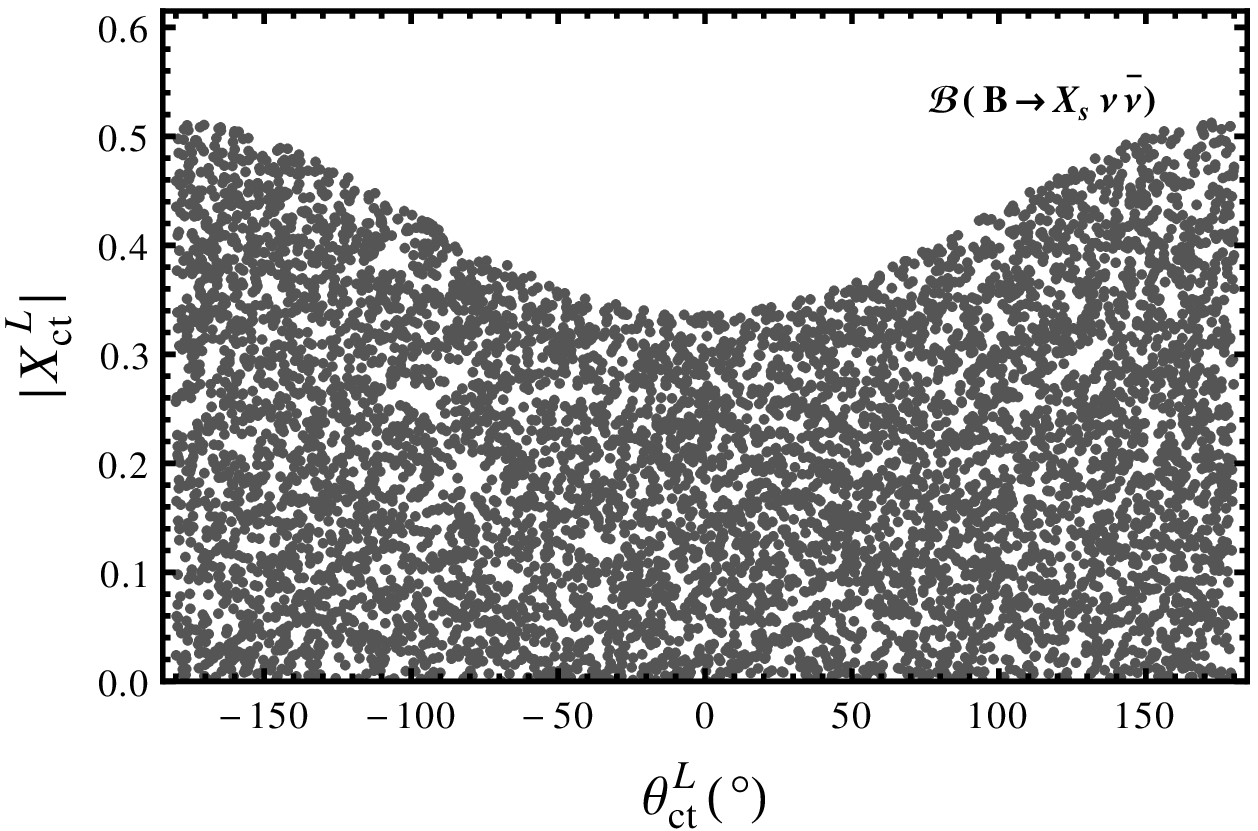}}\hspace{0.5in}
 \subfigure{\includegraphics [width=7cm]{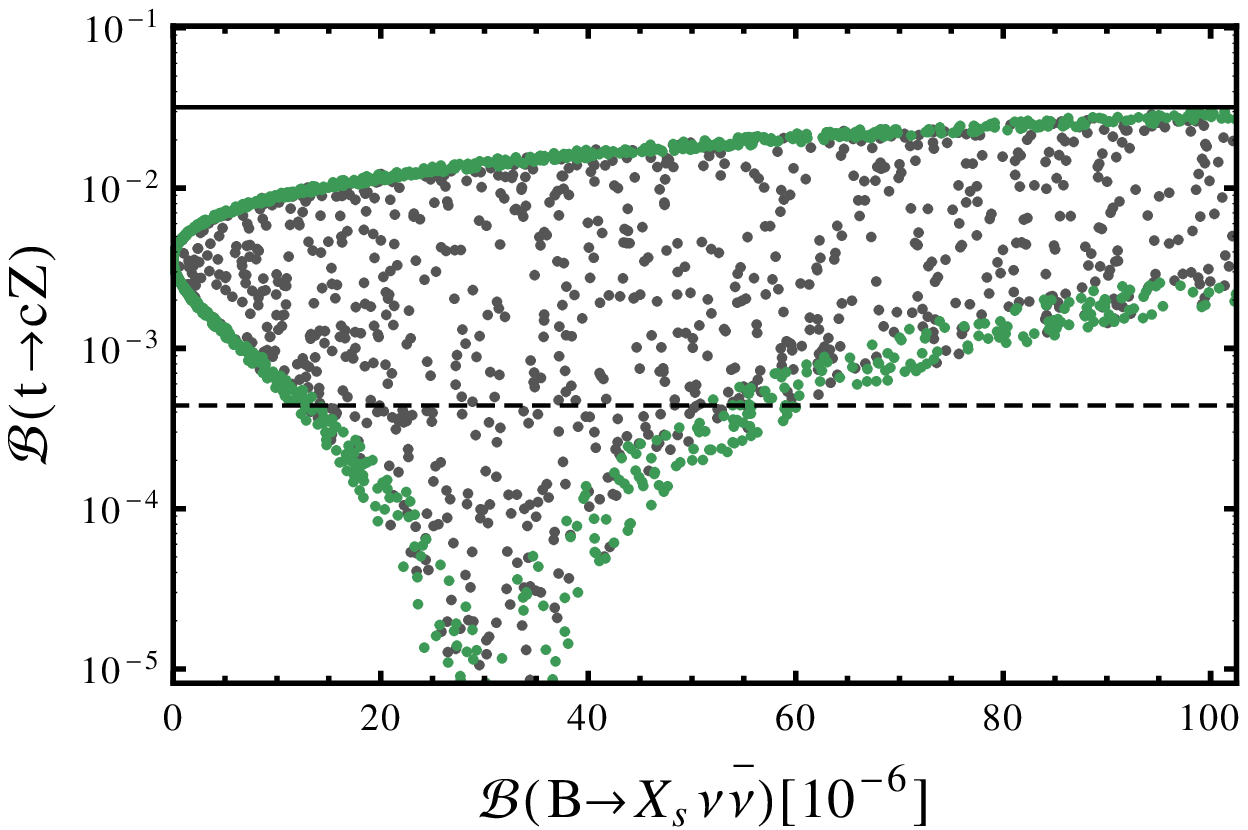}}
\caption{\small Upper bounds on the anomalous coupling $|X_{ct}^L|$ as a function of $\theta _{ct}^L$, and correlations between the rare B-meson and $t\to c Z$ decays. The allowed regions by rare B-meson decays are shown in dark and green points, with the latter obtained in the case of real coupling $X_{ct}^L$. The horizontal solid and dashed lines denote the D0 bound~\cite{Abazov:2011qf} and the ATLAS $5\sigma$ discovery potential at $L=10~{\rm fb}^{-1}$~\cite{Carvalho:2007yi}, respectively. The vertical solid line is the current experimental bounds on rare B-meson decays.}
\label{figure-Xct}
\end{figure}

For the decay $B_s\to \mu^+\mu^-$, the effect of anomalous $tcZ$ coupling results in a deviation of the function $Y(x)$ from its SM result, and the branching ratio can be formally written as
\begin{align}\label{bs2mumu_numerics}
\mathcal B(B_s\to \mu^+\mu^-) &\sim \Bigl|Y(x_t) + C_{0,b\to s}^{\rm NP}\Bigr|^2 \nonumber \\
&\sim \Bigl|0.96 + 16.91|X_{ct}^L|\,e^{i(\theta_{ct}^L+\beta_s)} - 0.04|X_{ct}^R|\,e^{i(\theta_{ct}^R+\beta_s)}\Bigr|^2\,,
\end{align}
where $\beta_s=-\arg(-\frac{V_{cs}V_{cb}^*}{V_{ts}V_{tb}^*})\simeq 1.04^\circ$, is the phase associated with the CKM matrix element $V_{ts}$. From Eq.~(\ref{bs2mumu_numerics}), one can see that, compared to the left-handed coupling, $X_{ct}^L=|X_{ct}^L|e^{i\theta_{ct}^L}$, the effect of right-handed coupling, $X_{ct}^R=|X_{ct}^R|e^{i\theta_{ct}^R}$, on the branching ratio is quite small, consistent with the observation made in Sec.~\ref{sec-theo}. Thus, we shall neglect the right-handed coupling $X_{ct}^R$ in the following discussions.

For a generic complex coupling $X_{ct}^L$, the branching ratio $\mathcal B(B_s\to \mu^+\mu^-)$ depends also on the phase $\theta_{ct}^L$. It can be seen from Eq.~(\ref{bs2mumu_numerics}) that, for a given value $|X_{ct}^L|$, the NP contribution is constructive to the SM one in the region $\theta_{ct}^L\approx -\beta_s$, whereas in the region $\theta_{ct}^L\approx 180^\circ-\beta_s$, the interference between them becomes destructive. This can be clearly seen from Fig.~\ref{figure-Xct}, where the upper bounds on the anomalous coupling $|X_{ct}^L|$ as a function of $\theta _{ct}^L$, as well as the correlations between rare B-meson and $t\to c Z$ decays are shown.

For the decays $B\to X_s\nu\bar\nu$, $B\to K \nu\bar\nu$ and $B\to K^* \nu\bar\nu$, which are all induced by the quark-level transition $b\to s\nu\bar\nu$, the effect of anomalous $tcZ$ coupling on the branching ratios is to replace the SM function $X(x_t)$ with $X(x_t) + C_{0,b\to s}^{\rm NP}$, and hence we have
\begin{align}\label{b2smumu_numerics}
\mathcal B(B\to K^{(*)} \nu\bar\nu, X_s\nu\bar\nu)
&\sim \Bigl|X(x_t) + C_{0,b\to s}^{\rm NP}\Bigr|^2 \nonumber \\
&\sim \Bigl|1.48 + 16.91|X_{ct}^L|\,e^{i(\theta_{ct}^L+\beta_s)} - 0.04|X_{ct}^R|\,e^{i(\theta_{ct}^R+\beta_s)}\Bigr|^2\,,
\end{align}
where the NP contribution is the same as discussed in $B_s\to \mu^+\mu^-$. However, due to the less stringent experimental bounds and the large theoretical uncertainties, the current constraints on the coupling $|X_{ct}^L|$ from these rare B-meson decays are still rather loose.

\begin{table}[t]
\begin{center}
\caption{\label{table-Xct} \small Bounds on the magnitude $|X_{ct}^L|$ from the purely leptonic $B_s\to\mu^+\mu^-$ decay, with some specific values of the phase $\theta_{ct}^L$. In the last row, we also give the corresponding predicted upper limit on $\mathcal B(t\to cZ)$.}
\vspace{0.2cm}
\doublerulesep 0.8pt \tabcolsep 0.20in
\begin{tabular}{l c c}
\hline\hline
& $\theta_{ct}^L=0^\circ$ & $\theta_{ct}^L=180^\circ$ \\
\hline
$\mathcal B(B_s\to \mu^+\mu^-)$ & $<0.043$ & $<0.16$ \\
D0 bound & $<0.26$ & $<0.26$ \\
\hline
$\mathcal B(t\to cZ)$ & $<8.4 \times 10^{-4}$ & $<0.011$ \\
\hline\hline
\end{tabular}
\end{center}
\end{table}

As is shown in Fig.~\ref{figure-Xct}, the potentially large top-quark anomalous coupling effect is reflected in the stringent bound on its magnitude $|X_{ct}^L|$, which is currently dominated by the purely leptonic $B_s\to\mu^+\mu^-$ decay. From the numerical results given in Table~\ref{table-Xct}, one can see that, with the bound from $\mathcal B(B_s\to\mu^+\mu^-)$ taken into account, the predicted upper limit of $\mathcal B (t\to cZ)$ is lower than the D0 bound~\cite{Abazov:2011qf}. This is also evident from the correlation plot between $\mathcal B(B_s\to \mu^+\mu^-)$ and $\mathcal B(t\to cZ)$ depicted in Fig.~\ref{figure-Xct}. In particular, the predicted $\mathcal B(t\to cZ)$ is about of the same order as the $5\sigma$ discovery potential of ATLAS with an integrated luminosity of $L=10~{\rm fb}^{-1}$~\cite{Carvalho:2007yi}.

\subsection{$B_d\to\mu^+\mu^-$ and $s\to d\nu\bar\nu$ decays with anomalous coupling $X_{ut}^L$}

\begin{figure}[t]
\centering
 \subfigure{\includegraphics[width=7cm]{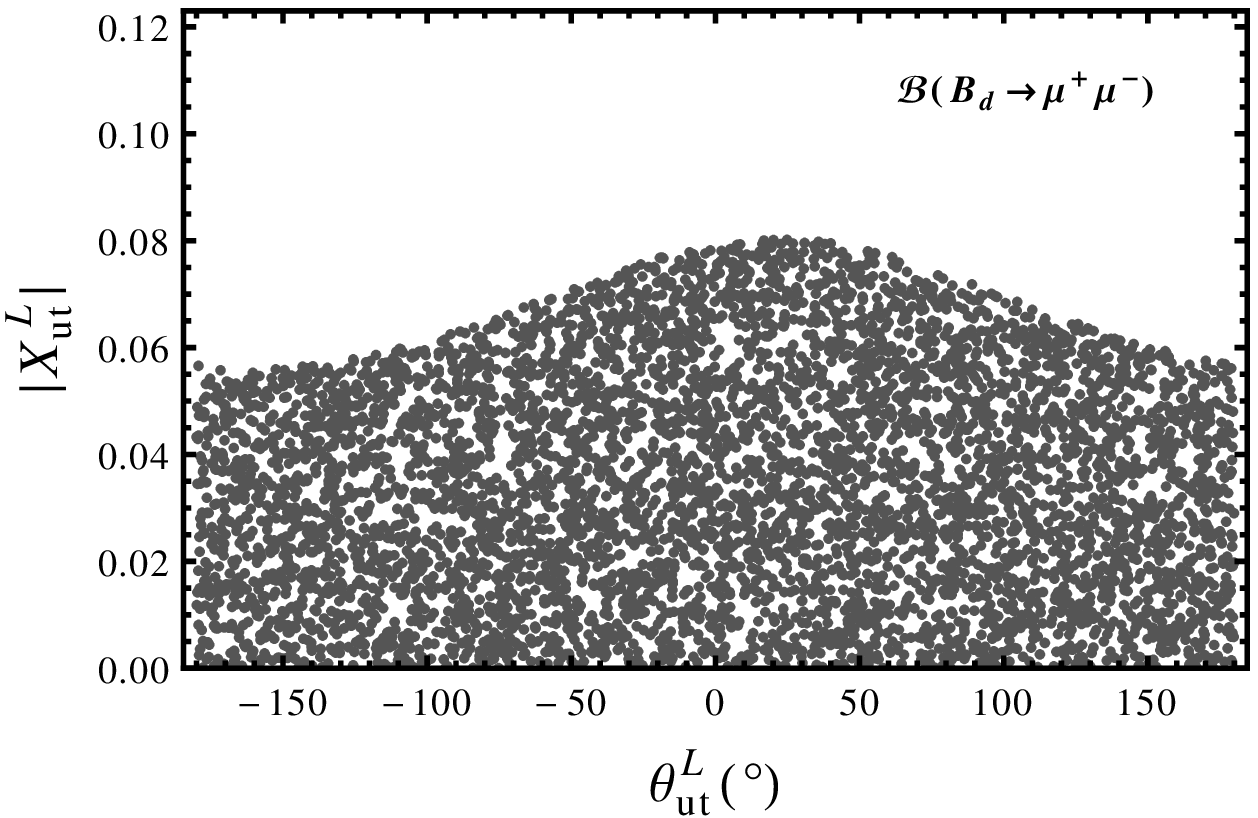}}\hspace{0.5in}
 \subfigure{\includegraphics[width=7cm]{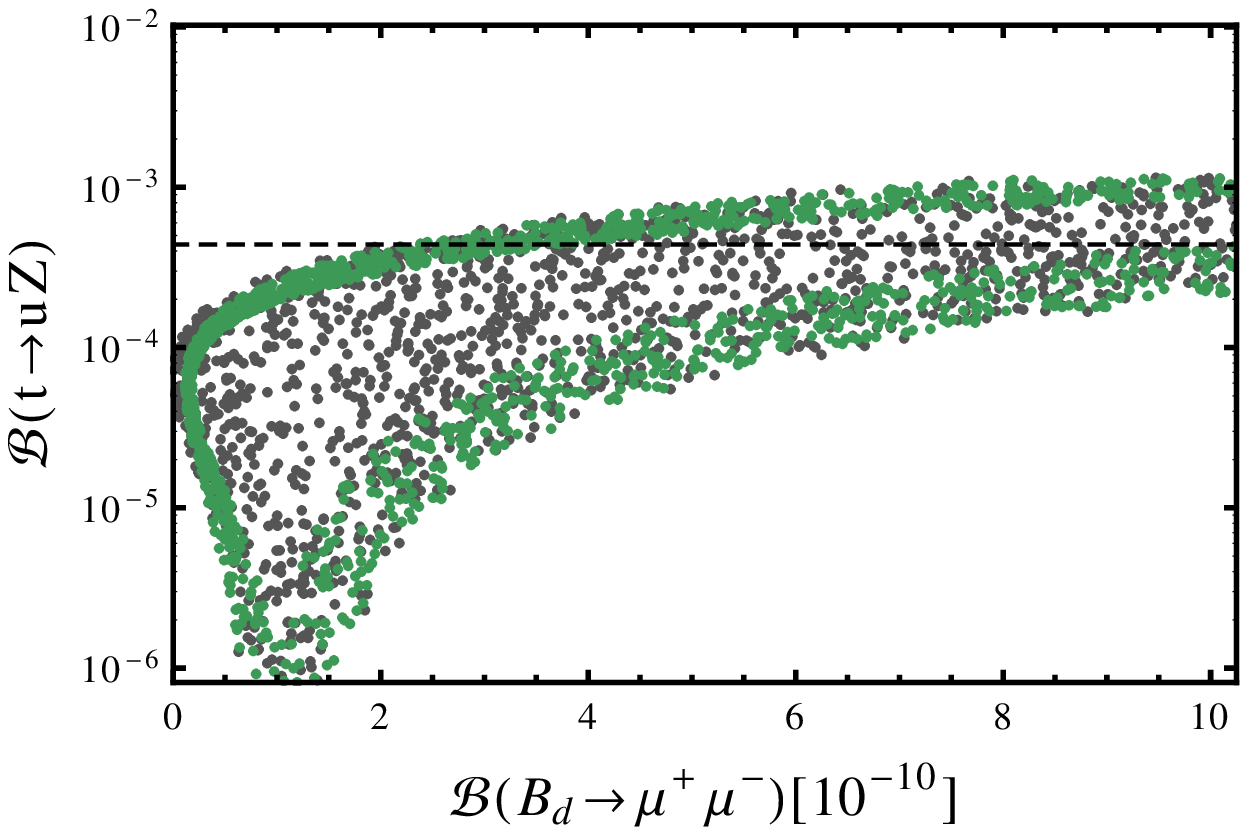}}
 \subfigure{\includegraphics[width=7cm]{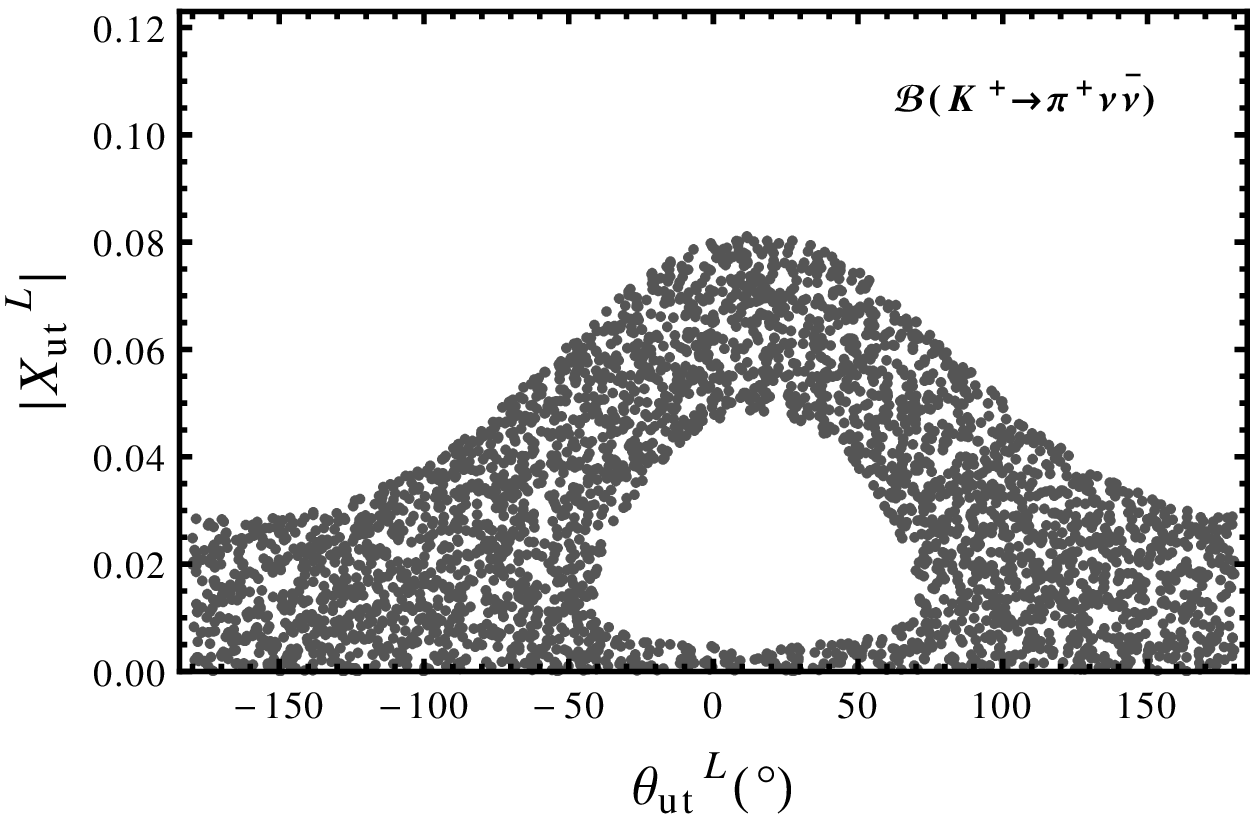}}\hspace{0.5in}
 \subfigure{\includegraphics[width=7cm]{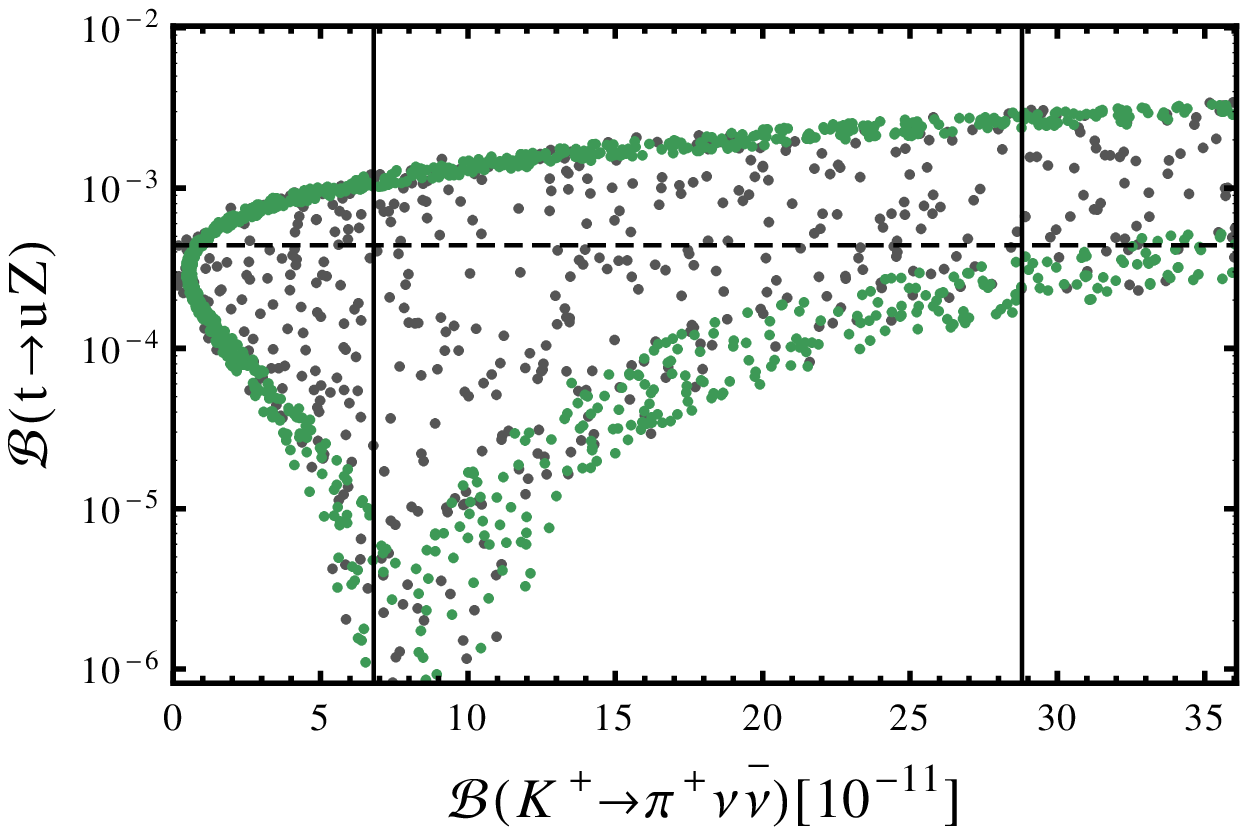}}
 \subfigure{\includegraphics[width=7cm]{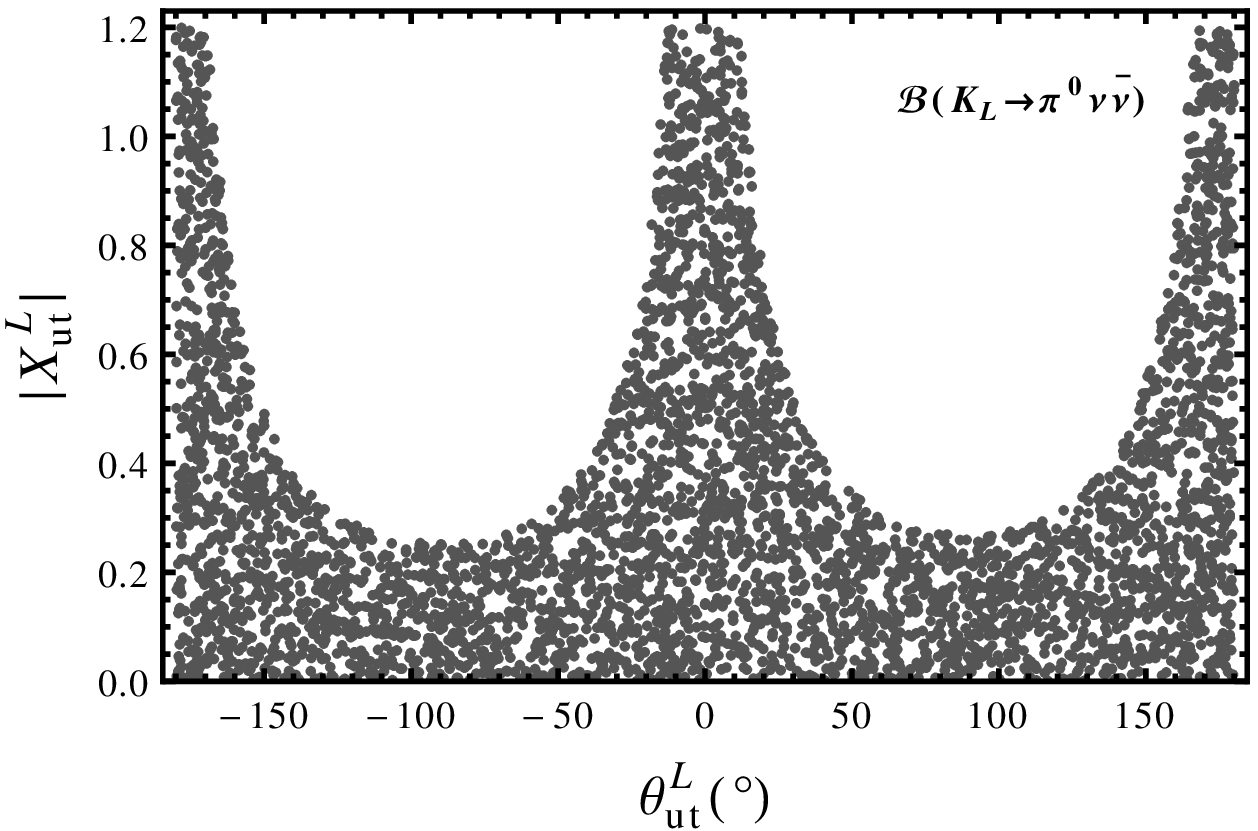}}\hspace{0.5in}
 \subfigure{\includegraphics[width=7cm]{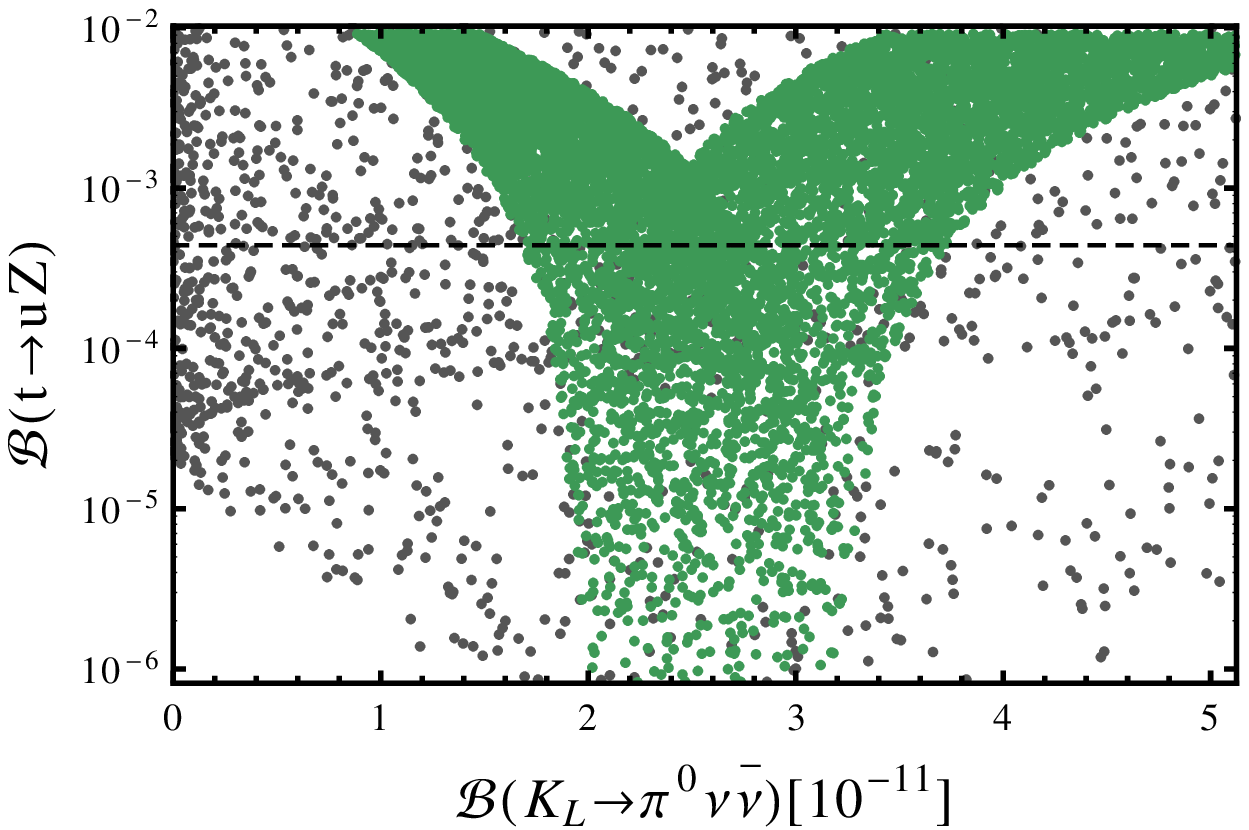}}
\caption{\small Upper bounds on the anomalous coupling $|X_{ut}^L|$ as a function of $\theta _{ut}^L$, and correlations between the rare B- and K-meson and the $t\to uZ$ decays. The other captions are the same as in Fig.~\ref{figure-Xct}.}
\label{figure-Xut}
\end{figure}

For the decay $B_d\to\mu^+\mu^-$, the anomalous $tuZ$ coupling contributes to the branching ratio as
\begin{align}
\mathcal B (B_d\to\mu^+\mu^-) &\sim \Bigl|Y(x_t) + C_{0,b\to d}^{\rm NP}\Bigr|^2 \nonumber \\
&\sim \Bigl|0.96 - 80.08|X_{ut}^L|\,e^{i(\theta_{ut}^L-\beta)}
+ 0.00067|X_{ut}^R|\,e^{i(\theta_{ut}^R-\beta)}\Bigr|^2\,,
\end{align}
where $\beta=\arg(-\frac{V_{cd}V_{cb}^*}{V_{td}V_{tb}^*})\simeq 21.78^\circ$, is the phase associated with the CKM matrix element $V_{td}$. The suppression of right-handed coupling is more evident and can be therefore neglected, since its contribution is accompanied by a much smaller factor $\sqrt x_u=m_u/m_W$, see Eq.~(\ref{C0b2d}). Here the interference between the SM and the NP contributions is destructive in the region $\theta_{ut}^L \approx \beta$, whereas constructive in the region $\theta_{ut}^L\approx \beta-180^\circ$. This is clearly shown in Fig.~\ref{figure-Xut}, where the upper bounds on the anomalous coupling $|X_{ut}^L|$ as a function of $\theta _{ut}^L$, as well as the correlations between rare B- and K-meson and $t\to u Z$ decays are depicted.

Although the current experimental bound on $\mathcal B(B_d\to\mu^+\mu^-)$ is still rather weak, it is interesting to note that, with its constraint on the coupling $X_{ut}^L$ taken into account, the predicted upper limit for $\mathcal B(t\to uZ)$ is lower than the D0 bound~\cite{Abazov:2011qf}. However, the limit is still comparable with the $5\sigma$ discovery potential of ATLAS with an integrated luminosity of $L=10~{\rm fb}^{-1}$~\cite{Carvalho:2007yi}, as shown in Fig.~\ref{figure-Xut}.

Due to the charm contribution, the effect of anomalous $tuZ$ coupling on the decay $K^+\to\pi^+\nu\bar\nu$ becomes somewhat complicated, and we have numerically
\begin{align}
\mathcal B(K^+\to\pi^+\nu\bar\nu) & \sim \Bigl|X(x_t) + C_{0,\bar{s}\to \bar{d}}^{\rm NP} + \frac{\lambda_c}{\lambda_t}\lambda^4(P_c+\delta P_{c,u})\Bigr|^2 \nonumber\\
& \sim \Bigl|1.48-80.08|X_{ut}^L|\,e^{-i(\theta_{ut}^L-\beta)}+0.68\,e^{i(\beta+\beta_s)}\Bigr|^2\,,
\end{align}
where $C_{0, \bar{s}\to \bar{d}}^{\rm NP}$ is the CP conjugation of $C_{0, s\to d}^{\rm NP}$ given by Eq.~(\ref{C0s2d}). Compared to the case of $B_d\to\mu^+\mu^-$, the constructive and destructive regions between the SM and the NP contributions are shifted a little bit by the charm sector, which can be seen by comparing the first two plots of the left column in Fig.~\ref{figure-Xut}.

Since the decay $K_L\to\pi^0\nu\bar\nu$ proceeds almost entirely through direct CP violation, only the imaginary part of the SM and the NP contributions affect its branching ratio. The final numerical result reads
\begin{align}
\mathcal B(K_L\to\pi^0\nu\bar\nu) &\sim \Bigl({\rm Im}\Bigl[\lambda_t X(x_t) + \lambda_t\,C_{0,\bar{s}\to \bar{d}}^{\rm NP}\Bigr]\Bigr)^2 \nonumber \\
&\sim \Bigl(1.48\sin (\beta+\beta_s)-80.08|X_{ut}^L|\sin(\theta_{ut}^L+\beta_s)\Bigr)^2.
\end{align}
In this case, the effect of anomalous $tuZ$ coupling depends strongly on its phase $\theta_{ut}^L$. In the regions $\theta_{ut}^L \approx -\beta_s$ or $\theta_{ut}^L \approx 180^\circ -\beta_s$, its effect is quite small and could even be zero. On the other hand, the largest effect comes from the region $\theta_{ut}^L \approx -90^\circ -\beta_s$, where the interference between the SM and the almost purely imaginary NP contributions are constructive. Consequently, the experimental bounds on the magnitude $|X_{ut}^L|$ depend crucially on the phase $\theta_{ut}^L$, as is shown in the last plot of the left column in Fig.~\ref{figure-Xut}.

From Fig.~\ref{figure-Xut}, we can see that, among all the three decay modes discussed here, the $K^+\to\pi^+\nu\bar\nu$ provides the most stringent constraint on the coupling $X_{ut}^L$. With the allowed values for $X_{ut}^L$, the predicted $\mathcal B(t\to uZ)$ is much lower than the D0 bound~\cite{Abazov:2011qf}. Furthermore, since the experimental measurement of $\mathcal B(K^+\to\pi^+\nu\bar\nu)$ is double-bounded, there are actually two solutions for the magnitude $|X_{ut}^L|$ in the destructive region $\theta_{ut}^L \approx \beta \approx 21.78^\circ$, with the larger one corresponding to the sign-flipped solution for the function $X(x_t)$.

Finally, for illustration, we give in Table~\ref{table-Xut} some numerical results for $|X_{ut}^L|$ with some specific values of the phase $\theta_{ut}^L$, where the scenarios S1 and S2 correspond to the same sign and sign-flipped solutions for the function $X(x_t)$, respectively. We can see that the predicted $\mathcal B(t\to uZ)$ is much lower than the D0 bound~\cite{Abazov:2011qf}, but is of the same order as the $5\sigma$ discovery potential of ATLAS with an integrated luminosity of $L=10~{\rm fb}^{-1}$~\cite{Carvalho:2007yi}.

\begin{table}[t]
\begin{center}
\caption{\label{table-Xut} \small Bounds on the magnitude $|X_{ut}^L|$ from the rare $K^+\to\pi^+\nu\bar\nu$ decay, with some specific values of the phase $\theta_{ut}^L$. The two solutions S1 and S2 correspond to the cases in which the sign of the function $X(x_t)$ is not flipped and flipped, respectively. In the last row, we also give the corresponding predicted upper limit on $\mathcal B(t\to cZ)$.}
\vspace{0.2cm}
\doublerulesep 0.8pt \tabcolsep 0.20in
\begin{tabular}{l c c c c c c }
\hline\hline
& $\theta_{ut}^L=0^\circ$~S1 & $\theta_{ut}^L=0^\circ$~S2 & $\theta_{ut}^L=180^\circ$ \\
\hline
$\mathcal B(K^+\to\pi^+\nu\bar\nu)$ & $<0.0048$ & $[0.047, 0.080]$ & $<0.029$ \\
D0 bound & $<0.26$ & $<0.26$ & $<0.26$ \\
\hline
$\mathcal B(t\to u Z)$ & $<1.0\times 10^{-5}$ & $[0.0010, 0.0030]$ & $<3.9\times 10^{-4}$ \\
\hline\hline
\end{tabular}
\end{center}
\end{table}

\section{Conclusions}
\label{sec-conclusion}

In this paper, we have studied the effects of anomalous $tqZ$ couplings on the rare B- and K-meson decays. Among the four operators in the effective Lagrangian given by Eq.~(\ref{Lagrangian-tqZ}), it is found that only the anomalous operator $\bar q \gamma^\mu P_{L} t Z_\mu$ could give a potentially large contribution to these rare decays. With the current experimental data on these decays, bounds on the coupling $X_{qt}^L$ of this operator are then derived. Our main conclusions are summarized as follows.

For the exclusive and inclusive $b\to s\nu\bar\nu$ decays, as well as the purely leptonic $B_s\to\mu^+\mu^-$ decays, we find that the main contribution comes from the anomalous $tcZ$ coupling, solely based on the countering of the associated CKM factors. On the other hand, the anomalous $tuZ$ coupling is found to dominate in the decays $B_d\to\mu^+\mu^-$, $K^+\to\pi^+\nu\bar\nu$ and $K_L\to\pi^0\nu\bar\nu$. Currently, the most stringent bounds on $tcZ$ and $tuZ$ couplings come from the decays $B_s\to\mu^+\mu^-$ and $K^+\to\pi^+\nu\bar\nu$, respectively.

For generical complex couplings $tcZ$ and $tuZ$, it is found that the interference between the SM and the NP contributions is constructive in the region $\theta_{ct}^L \approx -\beta_s$ for $B_s\to\mu^+\mu^-$ and $\theta_{ut}^L \approx \beta$ for $K^+\to\pi^+\nu\bar\nu$, respectively. Thus, the most stringent constraints on the strength of these anomalous couplings come from these regions.

From the correlations of the rare B- and K-meson decay with the rare $t\to qZ$ decays, we find that, with the constraints from the former taken into account, the predicted upper limit of $\mathcal B(t\to qZ)$ is lower than the D0 bound, but is still of the same order as the $5\sigma$ discovery potential of ATLAS with an integrated luminosity of $10~{\rm fb}^{-1}$.

Although the current experimental upper bounds on these rare decay processes are still rather weak, the measurements will be much improved at the LHCb, the future super-B factories, the NA62 at CERN, the KOTO at J-PARC, etc. Thus, closer correlations between the $t\to qZ$ and the rare B- and K-meson decays are expected in the near future, which will be very
helpful for the searches of  the anomalous top-quark FCNC decays at the LHC.

\section*{Acknowledgements}

The work was supported in part by the National Natural Science Foundation under contract Nos.~11075059, 11005032, 10735080 and 11047165. X.~Q. Li was also supported in part by MEC (Spain) under Grant FPA2007-60323 and by the Spanish Consolider Ingenio 2010 Programme CPAN (CSD2007-00042).

\begin{appendix}

\section*{Appendix A: Theoretical input parameters}
\label{app:input}

In this appendix, we collect all the relevant input parameters used in the numerical analysis of rare B- and K-meson decays, as well as the rare $t\to q Z$ decay.

\subsubsection*{The basic SM parameters}

First, we need some basic SM parameters, which are all taken from the Particle Data Group~\cite{Nakamura:2010zzi}
\begin{align}
& \alpha_s(m_Z)=0.1184\pm 0.0007, \quad \alpha(m_Z)=1/127.925, \quad G_F=1.16637\times 10^{-5}~{\rm GeV}^{-2}, \nonumber \\
& \sin^2\theta_W=0.23146, \quad m_W=80.399~{\rm GeV}, \quad m_Z=91.1876~{\rm GeV}, \nonumber \\
& m_{B^+}=5279.17~{\rm MeV}, \quad m_{B^0}=5279.50~{\rm MeV}, \quad m_{B_s}=5366.3~{\rm MeV},
\nonumber \\
& \tau_{B^+}=1.638~{\rm ps}, \quad \tau_{B^0}=1.525~{\rm ps}, \quad \tau_{B_s}=1.472~{\rm ps}.
\end{align}
We use two-loop running for $\alpha_s$ throughout this paper.

\subsubsection*{The CKM matrix elements}

For the CKM matrix elements, we adopt the Wolfenstein parametrization~\cite{Wolfenstein:1983yz} and choose the four parameters $A$, $\lambda$, $\rho$ and $\eta$ as fitted by the CKMfitter group~\cite{arXiv:1106.4041,Charles:2004jd}
\begin{equation}
A=0.816^{+0.011}_{-0.022}\,, \quad \lambda=0.22518^{+0.00036}_{-0.00077}\,, \quad \overline{\rho}=0.144^{+0.028}_{-0.019}\,, \quad \overline{\eta}=0.342^{+0.015}_{-0.014}\,,
\end{equation}
with $\overline{\rho}=\rho\,(1-\frac{\lambda^2}{2})$ and $\bar{\eta}=\eta\,(1-\frac{\lambda^2}{2})$.

\subsubsection*{The quark masses}

The quark masses given in different schemes are collected below
\begin{align}
 m_t^{\rm pole}&=173.2\pm 0.9~{\rm GeV}~\cite{arXiv:1107.5255}, \qquad m_b^{\rm 1S}=4.68\pm 0.03~{\rm GeV}~\cite{Bauer:2004ve}, \nonumber\\
 m_b(m_b)&=4.164\pm 0.023~{\rm GeV}~\cite{McNeile:2010ji}, \qquad m_c(m_c)=1.273\pm 0.006~{\rm GeV}~\cite{McNeile:2010ji}, \nonumber\\
 m_s(2~{\rm GeV})&=92.2\pm 1.3\,{\rm MeV}~\cite{McNeile:2010ji,Davies:2009ih}, \qquad m_u(2~{\rm GeV})=2.01\pm 0.14~{\rm MeV}~\cite{McNeile:2010ji,Davies:2009ih}.
\end{align}
To get the corresponding running quark masses at different scales, we use the NLO $\overline {\rm MS}$-on-shell conversion and running formulae collected, for example, in Ref.~\cite{Chetyrkin:2000yt}.

\subsubsection*{The nonperturbative hadronic parameters}

For $B_{s,d}\to \mu^+\mu^-$ decays, we need the B-meson decay constants, which are taken from~\cite{arXiv:0910.2928}
\begin{align}
f_{B_s}=238.8\pm 9.5~{\rm MeV} ,\qquad f_B=192.8\pm 9.9~{\rm MeV}.
\end{align}

For the $B\to K^{(*)}$ form factors appearing in $B\to K^{(*)}\nu\bar\nu$ decays, we adopt results obtained with the light-cone sum rule approach~\cite{hep-ph/0412079,hep-ph/0406232}
\begin{align}
V(q^2)&=\frac{r_1}{1-q^2/m_{R}^2}+\frac{r_2}{1-q^2/m_{\rm fit}^2}\,, &\;& \hspace{-0.7cm} {\rm with}\; r_1=0.923\,, r_2=-0.511\,, m_R=5.42\,, m_{\rm fit}^2=49.40\,,
\nonumber\\
A_1(q^2)&=\frac{r_2}{1-q^2/m_{\rm fit}^2}\,, && {\rm with}\; r_2=0.290\,, m_{\rm fit}^2=40.38\,, \nonumber\\
A_2(q^2)&=\frac{r_1}{1-q^2/m_{\rm fit}^2}+\frac{r_2}{(1-q^2/m_{\rm fit}^2)^2}\,, &\;& {\rm with}\; r_1=-0.084\,, r_2=0.342\,, m_{\rm fit}^2=52.00\,,
\nonumber\\
f_+^K(q^2)&=\frac{r_1}{1-q^2/m_{\rm fit}^2}+\frac{r_2}{(1-q^2/m_{\rm fit}^2)^2}\,, &\;& {\rm with}\; r_1=0.162\,, r_2=0.173\,, m_{\rm fit}^2=29.27\,,
\end{align}
which are valid in the full physical region, and the uncertainty is, to be conservative, assigned with the one at $q^2=0$, with $\delta V=\pm 0.033$, $\delta A_1=\pm 0.028$, $\delta A_2=\pm 0.027$, and $\delta f_+^K=\pm 0.041$.

\section*{Appendix B: The Inami-Lim functions X(x) and Y(x)}

The gauge-invariant functions $X(x)$ and $Y(x)$ appearing in rare B- and K-meson decays are given as~\cite{hep-ph/9901288,hep-ph/9901278,Inami:1980fz}
\begin{align}
X(x)=C(x)-4B(x)\,, \qquad Y(x)&=C(x)-B(x)\,,
\end{align}
where the basic Inami-Lim functions $C(x)$ and $B(x)$ correspond to the Z-penguin and the box diagram contribution, respectively. For convenience, their explicit expressions up to the NLO are given below~\cite{hep-ph/9901288,hep-ph/9901278,Inami:1980fz}
\begin{align}
Y(x)&=Y_0(x)+\frac{\alpha_s}{4\pi}Y_1(x),\nonumber\\[0.2cm]
Y_0(x)&=\frac{x}{8}\left[\frac{4-x}{1-x}+\frac{3x}{(1-x)^2}\ln x\right],\nonumber\\[0.2cm]
Y_1(x)&=\frac{10x+10x^2+4x^3}{3(1-x)^2}-\frac{2x-8x^2-x^3-x^4}{(1-x)^3}\ln x + \frac{2x-14x^2+x^3-x^4}{2(1-x)^3}\ln^2 x \nonumber\\
&\quad+\frac{2x+x^3}{(1-x)^2} L_2(1-x) + 8 x \frac{\partial Y_0(x)} {\partial x} \ln x_\mu\,,
\end{align}
and
\begin{align}
X(x)&=X_0(x)+\frac{\alpha_s}{4\pi}X_1(x),\nonumber\\[0.2cm]
X_0(x)&=\frac{x}{8}\left[-\frac{2+x}{1-x}+\frac{3x-6}{(1-x)^2}\ln x\right],\nonumber\\[0.2cm]
X_1(x)&=-\frac{29x-x^2-4x^3}{3(1-x)^2}-\frac{x+9x^2-x^3-x^4}{(1-x)^3}\ln x \nonumber\\
&\quad + \frac{8x+4x^2+x^3-x^4}{2(1-x)^3}\ln^2 x -\frac{4x-x^3}{(1-x)^2} L_2(1-x)+8 x \frac{\partial X_0(x)} {\partial x} \ln x_\mu\,,
\end{align}
with $x_\mu=\mu^2/m_W^2$, and $L_2(1-x)=\int_1^xdt \frac{\ln t}{1-t}$.

\end{appendix}


\begin{thebibliography}{100}

\bibitem{Glashow:1970gm}
  S.~L.~Glashow, J.~Iliopoulos, L.~Maiani,
  Phys.\ Rev.\  {\bf D2}, 1285-1292 (1970).

\bibitem{Eilam:1990zc}
  G.~Eilam, J.~L.~Hewett, A.~Soni,
  Phys.\ Rev.\  {\bf D44}, 1473-1484 (1991),
  Erratum-ibid D{\bf 59}, 039901 (1999).

\bibitem{AguilarSaavedra:2004wm}
  J.~A.~Aguilar-Saavedra,
  Acta Phys.\ Polon.\  B {\bf 35}, 2695 (2004)
  [arXiv:hep-ph/0409342].

\bibitem{Beneke:2000hk}
  M.~Beneke, I.~Efthymiopoulos, M.~L.~Mangano, J.~Womersley, A.~Ahmadov, G.~Azuelos, U.~Baur, A.~Belyaev {\it et al.},
  [hep-ph/0003033];
  W.~Bernreuther,
  J.\ Phys.\ G {\bf G35}, 083001 (2008)
  [arXiv:0805.1333 [hep-ph]]; and references therein.

\bibitem{Abazov:2011qf}
  V.~M.~Abazov {\it et al.}  [D0 Collaboration],
  Phys.\ Lett.\  B {\bf 701}, 313 (2011)
  [arXiv:1103.4574 [hep-ex]].

\bibitem{Carvalho:2007yi}
  J.~Carvalho {\it et al.} [ ATLAS Collaboration ],
  Eur.\ Phys.\ J.\  {\bf C52}, 999-1019 (2007)
  [arXiv:0712.1127 [hep-ex]];
  F.~M.~A.~Veloso,
  CERN-THESIS-2008-106.

\bibitem{Benucci:2008zz}
  L.~Benucci, A.~Kyriakis,
  Nucl.\ Phys.\ Proc.\ Suppl.\  {\bf 177-178}, 258-260 (2008).

\bibitem{FCNC-top}
  T.~Han, K.~Whisnant, B.~L.~Young, X.~Zhang,
  Phys.\ Rev.\  {\bf D55}, 7241-7248 (1997)
  [hep-ph/9603247];
  Phys.\ Lett.\  {\bf B385}, 311-316 (1996)
  [hep-ph/9606231];
  F.~Larios, M.~A.~Perez, C.~P.~Yuan,
  Phys.\ Lett.\  {\bf B457}, 334-340 (1999)
  [hep-ph/9903394];
  G.~Burdman, M.~C.~Gonzalez-Garcia, S.~F.~Novaes,
  Phys.\ Rev.\  {\bf D61}, 114016 (2000)
  [hep-ph/9906329];
  P.~J.~Fox, Z.~Ligeti, M.~Papucci, G.~Perez, M.~D.~Schwartz,
  Phys.\ Rev.\  {\bf D78}, 054008 (2008)
  [arXiv:0704.1482 [hep-ph]];
  J.~P.~Lee, K.~Y.~Lee,
  Phys.\ Rev.\  {\bf D78}, 056004 (2008)
  [arXiv:0806.1389 [hep-ph]];
  J.~Drobnak, S.~Fajfer and J.~F.~Kamenik,
  Phys.\ Lett.\ B\ {\bf 701}, 234  (2011)
  [arXiv:1102.4347 [hep-ph]];
  J.~Drobnak, S.~Fajfer and J.~F.~Kamenik,
  Nucl.\ Phys.\ B\ {\bf 855}, 82  (2012)
  [arXiv:1109.2357 [hep-ph]].

\bibitem{arXiv:0802.1413}
  B.~Grzadkowski and M.~Misiak,
  Phys.\ Rev.\ D\ {\bf 78}, 077501  (2008)
  [Erratum-ibid.\ D\ {\bf 84}, 059903  (2011)]
  [arXiv:0802.1413 [hep-ph]].

\bibitem{Han:1995pk}
  T.~Han, R.~D.~Peccei, X.~Zhang,
  Nucl.\ Phys.\  {\bf B454}, 527-540 (1995)
  [hep-ph/9506461].

\bibitem{Yuan:2010vk}
  X.~Yuan, Y.~Hao, Y.~Yang,
  Phys.\ Rev.\  {\bf D83}, 013004 (2011)
  [arXiv:1010.1912 [hep-ph]].

\bibitem{arXiv:1105.0364}
  X.~-Q.~Li, Y.~-D.~Yang and X.~-B.~Yuan,
  JHEP\ {\bf 1108}, 075  (2011)
  [arXiv:1105.0364 [hep-ph]].

\bibitem{Buras:2009if}
  A.~J.~Buras,
  PoS E {\bf PS-HEP2009}, 024 (2009)
  [arXiv:0910.1032 [hep-ph]].

\bibitem{Asner:2010qj}
  D.~Asner {\it et al.} [ Heavy Flavor Averaging Group Collaboration ],
  [arXiv:1010.1589 [hep-ex]], and online update at http://www.slac.stanford.edu/xorg/hfag.

\bibitem{Appelquist:1974tg}
  T.~Appelquist, J.~Carazzone,
  Phys.\ Rev.\  {\bf D11}, 2856 (1975).

\bibitem{Grzadkowski:2010es}
  B.~Grzadkowski, M.~Iskrzynski, M.~Misiak, J.~Rosiek,
  JHEP {\bf 1010}, 085 (2010)
  [arXiv:1008.4884 [hep-ph]].

\bibitem{Buchmuller:1985jz}
  W.~Buchmuller, D.~Wyler,
  Nucl.\ Phys.\  {\bf B268}, 621 (1986).

\bibitem{AguilarSaavedra:2008zc}
  J.~A.~Aguilar-Saavedra,
  Nucl.\ Phys.\  {\bf B812}, 181-204 (2009)
  [arXiv:0811.3842 [hep-ph]].

\bibitem{hep-ph/9901288}
  G.~Buchalla and A.~J.~Buras,
  Nucl.\ Phys.\ B\ {\bf 548}, 309  (1999)
  [hep-ph/9901288].

\bibitem{hep-ph/9901278}
  M.~Misiak and J.~Urban,
  Phys.\ Lett.\ B\ {\bf 451}, 161  (1999)
  [hep-ph/9901278].

\bibitem{Inami:1980fz}
  T.~Inami and C.~S.~Lim,
  Prog.\ Theor.\ Phys.\  {\bf 65}, 297 (1981)
  [Erratum-ibid.\  {\bf 65}, 1772 (1981)].

\bibitem{Altmannshofer:2009ma}
  W.~Altmannshofer, A.~J.~Buras, D.~M.~Straub and M.~Wick,
  JHEP {\bf 0904}, 022 (2009)
  [arXiv:0902.0160 [hep-ph]].

\bibitem{Bartsch:2009qp}
  M.~Bartsch, M.~Beylich, G.~Buchalla and D.~N.~Gao,
  JHEP {\bf 0911}, 011 (2009)
  [arXiv:0909.1512 [hep-ph]].

\bibitem{Colangelo:1996ay}
  P.~Colangelo, F.~De Fazio, P.~Santorelli and E.~Scrimieri,
  Phys.\ Lett.\  B {\bf 395}, 339 (1997)
  [arXiv:hep-ph/9610297].

\bibitem{Hoang:1998ng}
  A.~H.~Hoang, Z.~Ligeti and A.~V.~Manohar,
  Phys.\ Rev.\ Lett.\  {\bf 82}, 277 (1999)
  [arXiv:hep-ph/9809423];
  Phys.\ Rev.\  D {\bf 59}, 074017 (1999)
  [arXiv:hep-ph/9811239].

\bibitem{Bauer:2004ve}
  C.~W.~Bauer, Z.~Ligeti, M.~Luke, A.~V.~Manohar and M.~Trott,
  Phys.\ Rev.\  D {\bf 70}, 094017 (2004)
  [arXiv:hep-ph/0408002].

\bibitem{Buchalla:1993bv}
  G.~Buchalla and A.~J.~Buras,
  Nucl.\ Phys.\ B {\bf 400}, 225 (1993).

\bibitem{Grossman:1995gt}
  Y.~Grossman, Z.~Ligeti and E.~Nardi,
  Nucl.\ Phys.\  B {\bf 465}, 369 (1996)
  [Erratum-ibid.\  B {\bf 480}, 753 (1996)]
  [arXiv:hep-ph/9510378].

\bibitem{Falk:1995me}
  A.~F.~Falk, M.~E.~Luke and M.~J.~Savage,
  Phys.\ Rev.\ D {\bf 53}, 2491 (1996)
  [hep-ph/9507284].

\bibitem{hep-ph/0405132}
  A.~J.~Buras, F.~Schwab and S.~Uhlig,
  Rev.\ Mod.\ Phys.\ {\bf 80}, 965  (2008)
  [hep-ph/0405132].

\bibitem{arXiv:1107.6001}
  V.~Cirigliano, G.~Ecker, H.~Neufeld, A.~Pich and J.~Portoles,
  arXiv:1107.6001 [hep-ph].

\bibitem{hep-ph/9308272}
  G.~Buchalla and A.~J.~Buras,
  Nucl.\ Phys.\ B\ {\bf 412}, 106  (1994)
  [hep-ph/9308272];
  Phys.\ Rev.\ D\ {\bf 54}, 6782  (1996)
  [hep-ph/9607447].

\bibitem{CKM}
  N.~Cabibbo,
  Phys.\ Rev.\ Lett.\  {\bf 10}, 531-533 (1963);
  M.~Kobayashi, T.~Maskawa,
  Prog.\ Theor.\ Phys.\  {\bf 49}, 652-657 (1973).

\bibitem{arXiv:0705.2025}
  F.~Mescia and C.~Smith,
  Phys.\ Rev.\ D\ {\bf 76}, 034017  (2007)
  [arXiv:0705.2025 [hep-ph]].

\bibitem{428572}
  W.~J.~Marciano and Z.~Parsa,
  Phys.\ Rev.\ D\ {\bf 53}, 1  (1996).

\bibitem{hep-ph/0508165}
  A.~J.~Buras, M.~Gorbahn, U.~Haisch and U.~Nierste,
  Phys.\ Rev.\ Lett.\ {\bf 95}, 261805  (2005)
  [hep-ph/0508165];
  JHEP\ {\bf 0611}, 002  (2006)
  [hep-ph/0603079].

\bibitem{hep-ph/0503107}
  G.~Isidori, F.~Mescia and C.~Smith,
  Nucl.\ Phys.\ B\ {\bf 718}, 319  (2005)
  [hep-ph/0503107].

\bibitem{BNL-42227}
  L.~S.~Littenberg,
  Phys.\ Rev.\ D\ {\bf 39}, 3322  (1989).

\bibitem{Li:1990qf}
  C.~S.~Li, R.~J.~Oakes, T.~C.~Yuan,
  Phys.\ Rev.\  {\bf D43}, 3759-3762 (1991).

\bibitem{Zhang:2010bm}
  J.~J.~Zhang, C.~S.~Li, J.~Gao, H.~X.~Zhu, C.~-P.~Yuan and T.~-C.~Yuan,
  Phys.\ Rev.\ D {\bf 82}, 073005 (2010)
  [arXiv:1004.0898 [hep-ph]].

\bibitem{Zhang:2008yn}
  J.~J.~Zhang, C.~S.~Li, J.~Gao, H.~Zhang, Z.~Li, C.~-P.~Yuan and T.~-C.~Yuan,
  Phys.\ Rev.\ Lett.\  {\bf 102}, 072001 (2009)
  [arXiv:0810.3889 [hep-ph]].

\bibitem{Drobnak:2010by}
  J.~Drobnak, S.~Fajfer and J.~F.~Kamenik,
  Phys.\ Rev.\ D {\bf 82}, 073016 (2010)
  [arXiv:1007.2551 [hep-ph]].

\bibitem{Drobnak:2010wh}
  J.~Drobnak, S.~Fajfer and J.~F.~Kamenik,
  Phys.\ Rev.\ Lett.\  {\bf 104}, 252001 (2010)
  [arXiv:1004.0620 [hep-ph]].

\bibitem{CMSLHCb}
  The CMS and LHCb collaborations,
  LHCb-CONF-2011-047, CMS PAS BPH-11-019.

\bibitem{:2011zq}
  R. Aaij {\it et al.}  [LHCb Collaboration],
  arXiv:1112.1600 [Unknown].

\bibitem{hep-ex/0010022}
  R.~Barate {\it et al.} [ALEPH Collaboration],
  Eur.\ Phys.\ J.\ C\ {\bf 19}, 213  (2001)
  [hep-ex/0010022].

\bibitem{arXiv:1009.1529}
  P.~del Amo Sanchez {\it et al.} [BABAR Collaboration],
  Phys.\ Rev.\ D\ {\bf 82}, 112002  (2010)
  [arXiv:1009.1529 [hep-ex]].

\bibitem{arXiv:0808.1338}
  B.~Aubert {\it et al.} [BABAR Collaboration],
  Phys.\ Rev.\ D\ {\bf 78}, 072007  (2008)
  [arXiv:0808.1338 [hep-ex]].

\bibitem{arXiv:0808.2459}
  A.~V.~Artamonov {\it et al.} [E949 Collaboration],
  Phys.\ Rev.\ Lett.\ {\bf 101}, 191802  (2008)
  [arXiv:0808.2459 [hep-ex]];
  A.~V.~Artamonov {\it et al.} [BNL-E949 Collaboration],
  Phys.\ Rev.\ D\ {\bf 79}, 092004  (2009)
  [arXiv:0903.0030 [hep-ex]].

\bibitem{arXiv:0911.4789}
  J.~K.~Ahn {\it et al.} [E391a Collaboration],
  Phys.\ Rev.\ D\ {\bf 81}, 072004  (2010)
  [arXiv:0911.4789 [hep-ex]].

\bibitem{arXiv:1107.2304}
  T.~Aaltonen {\it et al.} [CDF Collaboration],
  Phys.\ Rev.\ Lett.\ {\bf 107}, 191801  (2011)
  [arXiv:1107.2304 [hep-ex]].

\bibitem{arXiv:1107.5834}
  S.~Chatrchyan {\it et al.} [CMS Collaboration],
  arXiv:1107.5834 [hep-ex].

\bibitem{arXiv:1110.2411}
  M.~-O.~Bettler,
  arXiv:1110.2411 [hep-ex];
  J.~Serrano,
  arXiv:1111.2620 [hep-ex].

\bibitem{hep-ex/0507034}
  K.~Abe {\it et al.} [Belle Collaboration],
  hep-ex/0507034.

\bibitem{arXiv:0707.0138}
  K.~-F.~Chen {\it et al.} [BELLE Collaboration],
  Phys.\ Rev.\ Lett.\ {\bf 99}, 221802  (2007)
  [arXiv:0707.0138 [hep-ex]].

\bibitem{Nakamura:2010zzi}
  KNakamura {\it et al.} [ Particle Data Group Collaboration ],
  J.\ Phys.\ G {\bf G37}, 075021 (2010).

\bibitem{Wolfenstein:1983yz}
  L.~Wolfenstein,
  Phys.\ Rev.\ Lett.\  {\bf 51}, 1945 (1983).

\bibitem{arXiv:1106.4041}
  J.~Charles, O.~Deschamps, S.~Descotes-Genon, R.~Itoh, H.~Lacker, A.~Menzel, S.~Monteil and V.~Niess {\it et al.},
  Phys.\ Rev.\ D\ {\bf 84}, 033005  (2011)
  [arXiv:1106.4041 [hep-ph]].

\bibitem{Charles:2004jd}
  J.~Charles {\it et al.} [ CKMfitter Group Collaboration ],
  Eur.\ Phys.\ J.\  {\bf C41}, 1-131 (2005)
  [hep-ph/0406184], and updated at http://ckmfitter.in2p3.fr/.

\bibitem{arXiv:1107.5255}
  M.~Lancaster [Tevatron Electroweak Working Group and for the CDF and D0 Collaborations],
  arXiv:1107.5255 [hep-ex].

\bibitem{McNeile:2010ji}
  C.~McNeile, C.~T.~H.~Davies, E.~Follana, K.~Hornbostel, G.~P.~Lepage,
  Phys.\ Rev.\  {\bf D82}, 034512 (2010)
  [arXiv:1004.4285 [hep-lat]].

\bibitem{Davies:2009ih}
  C.~T.~H.~Davies, C.~McNeile, K.~Y.~Wong, E.~Follana, R.~Horgan, K.~Hornbostel, G.~P.~Lepage, J.~Shigemitsu {\it et al.},
  Phys.\ Rev.\ Lett.\  {\bf 104}, 132003 (2010)
  [arXiv:0910.3102 [hep-ph]].

\bibitem{Chetyrkin:2000yt}
  K.~G.~Chetyrkin, J.~H.~Kuhn, M.~Steinhauser,
  Comput.\ Phys.\ Commun.\  {\bf 133}, 43-65 (2000)
  [hep-ph/0004189].

\bibitem{arXiv:0910.2928}
  J.~Laiho, E.~Lunghi and R.~S.~Van de Water,
  Phys.\ Rev.\ D\ {\bf 81}, 034503  (2010)
  [arXiv:0910.2928 [hep-ph]].

\bibitem{hep-ph/0412079}
  P.~Ball and R.~Zwicky,
  Phys.\ Rev.\ D\ {\bf 71}, 014029  (2005)
  [hep-ph/0412079].

\bibitem{hep-ph/0406232}
  P.~Ball and R.~Zwicky,
  Phys.\ Rev.\ D\ {\bf 71}, 014015  (2005)
  [hep-ph/0406232].

\end{thebibliography}
\end{document}